\documentclass[a4paper,USenglish,cleveref,autoref,numberwithinsect]{lipics-v2021}

\usepackage{tikz,xcolor}
\title{A Homotopic Kuratowski Theorem}

\author{Lo{\"i}c Dubois}{University of Notre Dame \and \url{https://loic-dubois.github.io/}}{ldubois@nd.edu}{https://orcid.org/0000-0002-0366-4959}{}
\authorrunning{L. Dubois}
\Copyright{Lo{\"i}c Dubois}


\nolinenumbers 

\hideLIPIcs

\EventEditors{John Q. Open and Joan R. Access}
\EventNoEds{2}
\EventLongTitle{42nd Conference on Very Important Topics (CVIT 2016)}
\EventShortTitle{CVIT 2016}
\EventAcronym{CVIT}
\EventYear{2016}
\EventDate{December 24--27, 2016}
\EventLocation{Little Whinging, United Kingdom}
\EventLogo{}
\SeriesVolume{42}
\ArticleNo{23}

\begin{document}

\maketitle

\begin{abstract}
We prove that if a map from a graph $G$ to a compact orientable surface, possibly with boundary, cannot be deformed so as to remove all crossings, then neither can its restriction to some subgraph of $G$ with at most $72$ vertices of degree other than two. On the sphere, this is Kuratowski's theorem; in contrast, minimal obstructions to embeddability in a surface grow with its genus. Our proof reduces the problem to planarity in a disk with order constraints, and identifies a condition under which this problem has small obstructions. Our main tool is a homotopic version of depth-first search trees, which underlie classical planarity tests and the extraction of Kuratowski subgraphs.
\end{abstract}

\section{Introduction}\label{sec:intro}

In this paper, we ask when a map $f$ from a graph $G$ to a surface $S$ \emph{can be untangled}: when it can be deformed so as to remove all crossings, in other words, when it is homotopic to an injective map. A graph vertex is \emph{topological} if it does not have degree two. We prove:

\begin{theorem}\label{thm:main}
Let $S$ be a compact orientable surface, possibly with boundary, $G$ a graph, and $f:G\to S$ a map. If $f$ cannot be untangled, then neither can $f|_{G'}$ for some $G' \subseteq G$ with at most $72$ topological vertices.
\end{theorem}

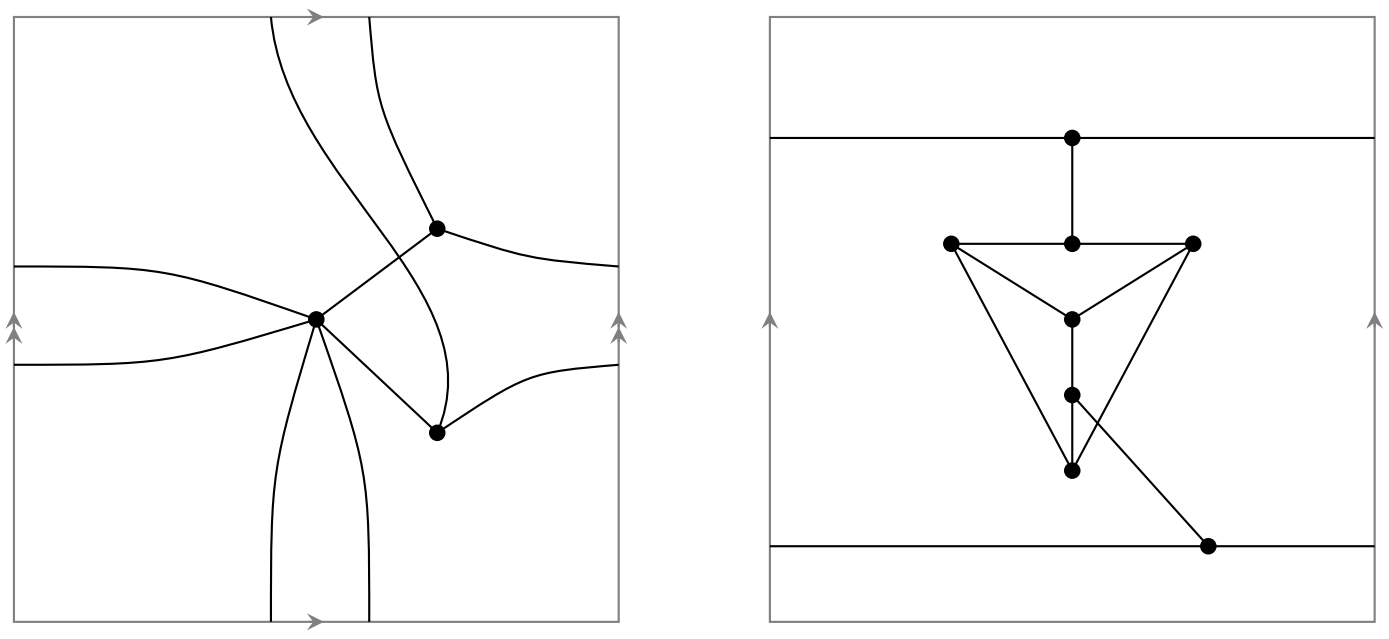
\begin{figure}[ht]
\centering
\begin{tikzpicture}[
  vtx/.style={circle,fill,inner sep=1.1pt},
  sec/.style={},
  >=stealth]
  \begin{scope}
  \draw[gray] (-2,-2) rectangle (2,2);
  \draw[gray,->] (-0.05,2)--(0.05,2);   \draw[gray,->] (-0.05,-2)--(0.05,-2);
  \draw[gray,->] (2,-0.05)--(2,0.05);   \draw[gray,->] (-2,-0.05)--(-2,0.05);
  \draw[gray,->] (2,-0.15)--(2,-0.05);  \draw[gray,->] (-2,-0.15)--(-2,-0.05);
  \coordinate (o) at (0,0); \coordinate (u) at (0.8,0.6); \coordinate (v) at (0.8,-0.75);
  \draw (o)--(u);
  \draw (o) .. controls (-1,0.35) .. (-2,0.35);  \draw (2,0.35) .. controls (1.4,0.4) .. (u);
  \draw (o) .. controls (0.35,-1) .. (0.35,-2);  \draw (0.35,2) .. controls (0.4,1.4) .. (u);
  \draw[sec] (o)--(v);
  \draw[sec] (o) .. controls (-1,-0.3) .. (-2,-0.3);  \draw[sec] (2,-0.3) .. controls (1.4,-0.35) .. (v);
  \draw[sec] (o) .. controls (-0.3,-1) .. (-0.3,-2);  \draw[sec] (-0.3,2) .. controls (-0.2,0.9) and (1.2,0.2) .. (v);
  \foreach \p in {o,u,v} \node[vtx] at (\p) {};
  \end{scope}
  \begin{scope}[xshift=5cm]
  \draw[gray] (-2,-2) rectangle (2,2);
  \draw[gray,->] (2,-0.05)--(2,0.05);   \draw[gray,->] (-2,-0.05)--(-2,0.05);
  \coordinate (1) at (-0.8,0.5); \coordinate (2) at (0.8,0.5);
  \coordinate (3) at (0,-1);     \coordinate (4) at (0,0);
  \coordinate (a) at (0,0.5);    \coordinate (b) at (0,-0.5);
  \draw (1)--(2)--(3)--(1) (1)--(4)--(2) (4)--(3);
  \coordinate (a') at (0,1.2);   \coordinate (b') at (0.9,-1.5);
  \draw (a)--(a') (b)--(b');
  \draw (-2,1.2) -- (2,1.2);
  \draw (-2,-1.5) -- (2,-1.5);
  \foreach \p in {1,2,3,4,a,b,a',b'} \node[vtx] at (\p) {};
  \end{scope}
\end{tikzpicture}
\caption{Minimal maps to the torus and the annulus that cannot be untangled.}
\label{fig:minimal-maps}
\end{figure}

Remarkably, this bound does not grow with the genus of $S$, in sharp contrast with obstructions to embeddability. Theorem~\ref{thm:main} also yields a polynomial-time algorithm for computing a small certificate that $f$ cannot be untangled: repeatedly delete an edge or isolated vertex of $G$ while the restriction of $f$ still cannot be untangled, testing each step with an algorithm of Colin de Verdi\`ere, Despr\'e, and Dubois~\cite{colin2026untangling}. Whether these results extend to non-orientable surfaces remains open, as do the optimal constant, between $8$ (Figure~\ref{fig:minimal-maps}, right) and $72$, and the graphs involved in minimal maps. We fix a compact orientable surface $S$.

\subsection{Related works}

When $S$ is the sphere, all maps from $G$ to $S$ are homotopic, so $f$ can be untangled if and only if $G$ is planar, and Theorem~\ref{thm:main} reduces to Kuratowski's theorem~\cite{Kuratowski1930}. Without homotopy constraints, embeddability in a fixed surface is characterized by finitely many minimal forbidden subgraphs~\cite{RobertsonSeymour1990GM8}, but these are known only for the plane and the projective plane, which already has $103$ of them~\cite{GloverHunekeWang1979,Archdeacon1981}, and their number grows with the genus $g$ of the surface: $g+1$ disjoint copies of $K_{3,3}$ do not embed, while each proper subgraph does. 

When $G$ is a cycle, the question is whether a closed curve is homotopic to a simple one. It goes back to Poincar\'e~\cite{Poincare1904}, and was solved by Reinhart~\cite{Reinhart1962}, Chillingworth~\cite{Chillingworth1969,Chillingworth1972}, and Birman and Series~\cite{BirmanSeries1984}; Despr\'e and Lazarus gave a near-linear algorithm~\cite{DespreLazarus2019}. As shortest closed geodesics minimize intersections~\cite{FreedmanHassScott1982}, a family of closed curves can be untangled if and only if each curve and each pair of curves can be. 

Colin de Verdi\`ere, Despr\'e, and Dubois~\cite{colin2026untangling} decide in polynomial time whether a map can be untangled, using the recognition of weak embeddings~\cite{FulekKyncl2018HTApprox,AkitayaFulekToth2019}. Isotopy of embeddings was characterized by Ladegaillerie~\cite{Ladegaillerie1984} and decided efficiently by Colin de Verdi\`ere and de Mesmay~\cite{CdVdM2014Isotopy}. In the plane, embedded minors well-quasi-order plane graphs, so every minor-closed property has finitely many excluded minors, as proved by Medina, Mohar, and Salazar~\cite{medina2019plane}, and by Lunel and Maria~\cite{lunel2025wqo}.

Our problem is related to planarity in a disk where matched pairs of boundary vertices must have reversed edge orders, a special case of synchronized planarity, which Bl\"asius, Fink, and Rutter solve in quadratic time~\cite{BlasiusFinkRutter2023}. That algorithm, like Fulek and Kyn\v{c}l's Hanani--Tutte-type characterization of weak embeddings~\cite{FulekKyncl2018HTApprox}, does not bound the size of obstructions.

Our proof relies on depth-first spanning trees, also called Tr\'emaux trees, which underlie planarity testing~\cite{HopcroftTarjan1974,deFraysseixRosenstiehl1985} and the extraction of Kuratowski subgraphs~\cite{Williamson1984}.

\subsection{Overview}

A \emph{loop graph} is a graph whose connected components each have a single vertex. Arguments of Colin de Verdi\`ere, Despr\'e, and Dubois~\cite{colin2026untangling} give:

\begin{lemma}[Appendix~\ref{app:loops}]\label{lem:loops}
Let $L$ be a loop graph and $\lambda : L \to S$. If $\lambda$ cannot be untangled, then $L$ has a subgraph $L'$ of at most four loops such that $\lambda|_{L'}$ cannot be untangled.
\end{lemma}

The same authors reduce general graphs to loop graphs~\cite{colin2026untangling}. Given a graph $G$, and a map $f:G\to S$, in each component of $G$, contract the image of a spanning tree to a point, bundle homotopic loops, and contract contractible ones. This turns $f$ into a composed map $\lambda\circ\mu$, with $L$ a loop graph, $\mu:G\to L$ and $\lambda:L\to S$: the \emph{factorization} of $f$ \emph{along} the spanning forest, defined up to homotopy of $\lambda$. A \emph{weak embedding} is the uniform limit of a sequence of embeddings.

\begin{lemma}[{\cite[Lemma~5.5]{colin2026untangling}}]\label{lem:factorization-untangle}
Let $\lambda\circ\mu$ be the factorization of $f:G\to S$ along a spanning forest of $G$. If $\lambda$ is an embedding, and if $f$ can be untangled, then $\mu$ is a weak embedding.
\end{lemma}

\begin{figure}[ht]
\centering
\begin{tikzpicture}[
  vtx/.style={circle,fill,inner sep=1.1pt},
  cvtx/.style={circle,draw,fill=white,inner sep=1.1pt},
  half/.style={gray},
  lbl/.style={font=\scriptsize},
  >=stealth]
\begin{scope}
  \fill[gray!12] (0,0) circle (0.95);
  \fill[gray!12] (-2,-0.5) rectangle (2,0.55);
  \fill[gray!12] (-0.55,-2) rectangle (0.35,2);
  \draw[gray] (-2,-2) rectangle (2,2);
  \draw[gray,->] (-0.05,2)--(0.05,2);   \draw[gray,->] (-0.05,-2)--(0.05,-2);
  \draw[gray,->] (2,-0.05)--(2,0.05);   \draw[gray,->] (-2,-0.05)--(-2,0.05);
  \draw[gray,->] (2,-0.15)--(2,-0.05);  \draw[gray,->] (-2,-0.15)--(-2,-0.05);
  \coordinate (u) at (0.35,0.25); \coordinate (v) at (-0.4,0.2); \coordinate (w) at (0.05,-0.4);
  \draw (u)--(v)--(w);
  \draw (u) .. controls (1,0.35) .. (2,0.38);  \draw (-2,0.38) .. controls (-1,0.38) .. (v);
  \draw (w) .. controls (1,-0.2) .. (2,-0.15); \draw (-2,-0.15) .. controls (-1,-0.1) .. (v);
  \draw (v) .. controls (-0.15,1) .. (-0.1,2); \draw (-0.1,-2) .. controls (-0.1,-1) .. (w);
  \foreach \p in {u,v,w} \node[vtx] at (\p) {};
  \node[lbl,above right] at (u) {$u$}; \node[lbl,above left] at (v) {$v$}; \node[lbl,below right] at (w) {$w$};
  \node[lbl] at (1.75,0.62) {$1$}; \node[lbl] at (1.75,-0.4) {$2$};
  \node[lbl] at (-1.75,0.62) {$1$}; \node[lbl] at (-1.75,-0.4) {$2$};
  \node at (0,-2.5) {(a)};
\end{scope}
\begin{scope}[xshift=5.2cm]
  \draw (0,0) circle (1.7);
  \coordinate (x) at (1.7,0); \coordinate (xb) at (-1.7,0);
  \coordinate (y) at (0,1.7); \coordinate (yb) at (0,-1.7);
  \coordinate (u) at (0.5,0.4); \coordinate (v) at (-0.55,0.3); \coordinate (w) at (0.1,-0.6);
  \draw (u)--(v)--(w);
  \draw[half] (u) .. controls (1.1,0.35) .. (x);
  \draw[half] (w) .. controls (1.1,-0.35) .. (x);
  \draw[half] (v) .. controls (-1.1,0.4) .. (xb);
  \draw[half] (v) .. controls (-1.1,-0.05) .. (xb);
  \draw[half] (v) .. controls (-0.2,1.2) .. (y);
  \draw[half] (w) .. controls (0,-1.2) .. (yb);
  \foreach \p in {u,v,w} \node[vtx] at (\p) {};
  \foreach \p in {x,xb,y,yb} \node[cvtx] at (\p) {};
  \node[lbl,above right] at (u) {$u$}; \node[lbl,above left] at (v) {$v$}; \node[lbl,below right] at (w) {$w$};
  \node[lbl,right] at (x) {$x$}; \node[lbl,left] at (xb) {$\bar x$};
  \node[lbl,above] at (y) {$y$}; \node[lbl,below] at (yb) {$\bar y$};
  \node[lbl] at (1.38,0.32) {$1$}; \node[lbl] at (1.38,-0.32) {$2$};
  \node[lbl] at (-1.38,0.38) {$1$}; \node[lbl] at (-1.38,-0.2) {$2$};
  \node at (0,-2.5) {(b)};
\end{scope}
\end{tikzpicture}
\caption{A drawing near a loop graph on the torus (a), and its gadget (b).}
\label{fig:gadget}
\end{figure}
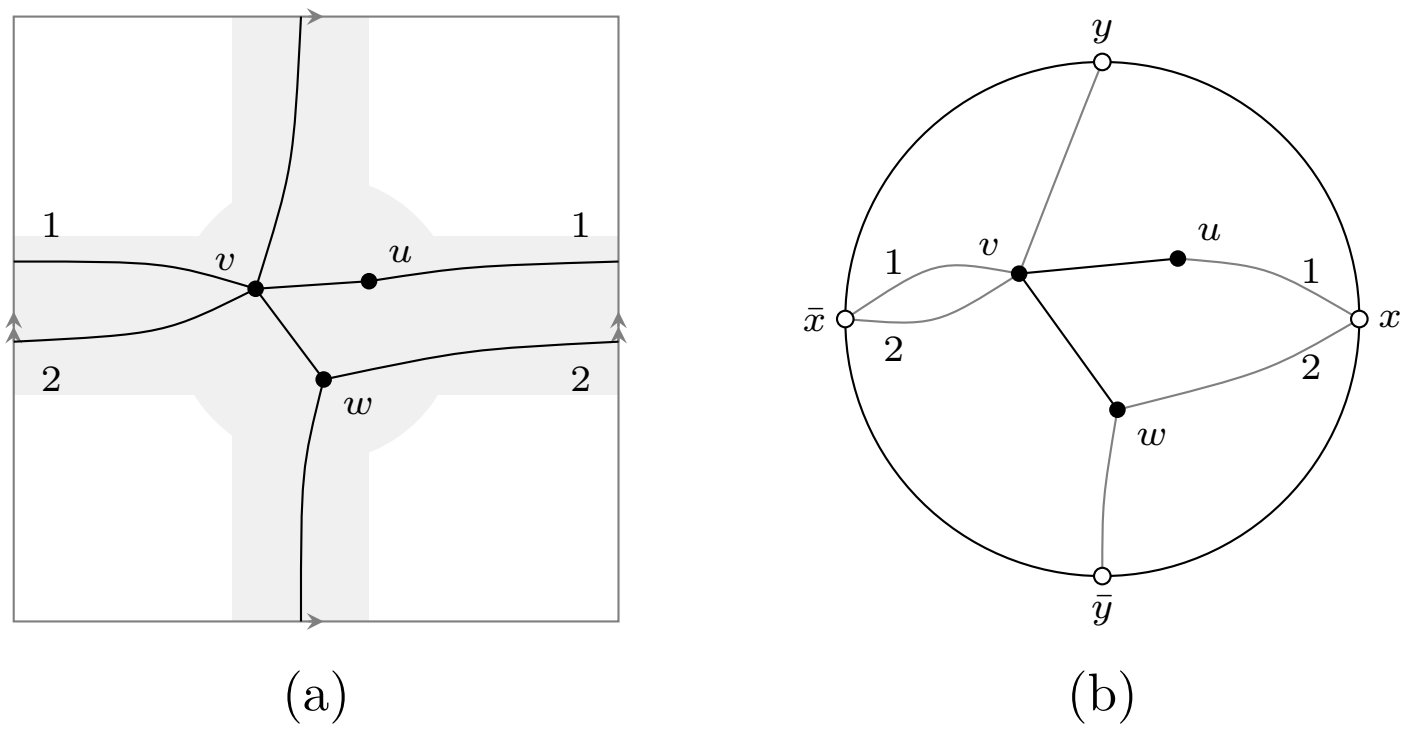

So, when $\lambda$ is an embedding, untangling $f$ amounts to perturbing $\mu$ into an embedding near $L$, for each component of $L$. We reformulate this as a planarity problem with order constraints. The embedding $\lambda$ orders the half-edges of $L$ cyclically around its vertex. We construct a cycle $C$ with one vertex per half-edge, in this order, cut each edge of $G$ sent around a loop into two matched halves, and attach them to the corresponding vertices of $C$ (Figure~\ref{fig:gadget}).

\begin{definition}
A \emph{gadget} $(H,C)$ is a cycle $C$ and a graph $H$ sharing no edge with $C$. The edges of $H$ meeting $C$ are \emph{halves}. They, and the vertices of $C$, are partitioned into \emph{matched} pairs, with matched halves at matched vertices, dropped from the notation for convenience.
\end{definition}

We only consider subgraphs $H'$ of $H$ containing the match of each of their halves. Such $H'$ is \emph{realizable} if $H'\cup C$ has a plane embedding with outer face $C$ that reverses the orders of the halves at matched vertices. Its \emph{joined graph} is obtained by joining matched halves into single edges and deleting $V(C)$, and its \emph{rank} is the cycle rank of this graph. For the gadget of a factorization, the joined graph of $H'$ \emph{corresponds} to a subgraph of $G$.

\begin{lemma}[Appendix~\ref{app:gadget}]\label{lem:gadget}
Let $H'\subseteq H$ correspond to $G' \subseteq G$. Assume that $H'-V(C)$ is connected. Then $f|_{G'}$ can be untangled if and only if $H'$ is realizable.
\end{lemma}

It is not clear that non-realizable gadgets have small non-realizable subgraphs: a seemingly related problem has arbitrarily large minimal obstructions~\cite[Section~5]{Mohar}. Our key contribution is a condition under which they do. A set $U\subseteq V(H)\setminus V(C)$ is \emph{rich} if the halves at $U$ reach at least two vertices of $C$. A spanning tree $T$ of $H-V(C)$, rooted at a vertex $r$, is \emph{grounded} if $V(T)$ is rich and, whenever a component $K$ of $T-r$ is rich, so is $V(T)\setminus K$. The rooted tree $(T,r)$ is \emph{depth-first} in a graph spanned by $T$ if every edge of that graph joins a vertex to itself or to one of its ancestors in $(T,r)$.

\begin{proposition}[Section~\ref{sec:core}]\label{prop:gadget}
Let $(H,C)$ be a gadget. Assume that $H-V(C)$ has a rooted spanning tree $(T,r)$ that is grounded and depth-first in the joined graph of $H$. If $H$ is not realizable, then neither is some $H'\subseteq H$ of rank at most $37$ such that $H'-V(C)$ is connected.
\end{proposition}

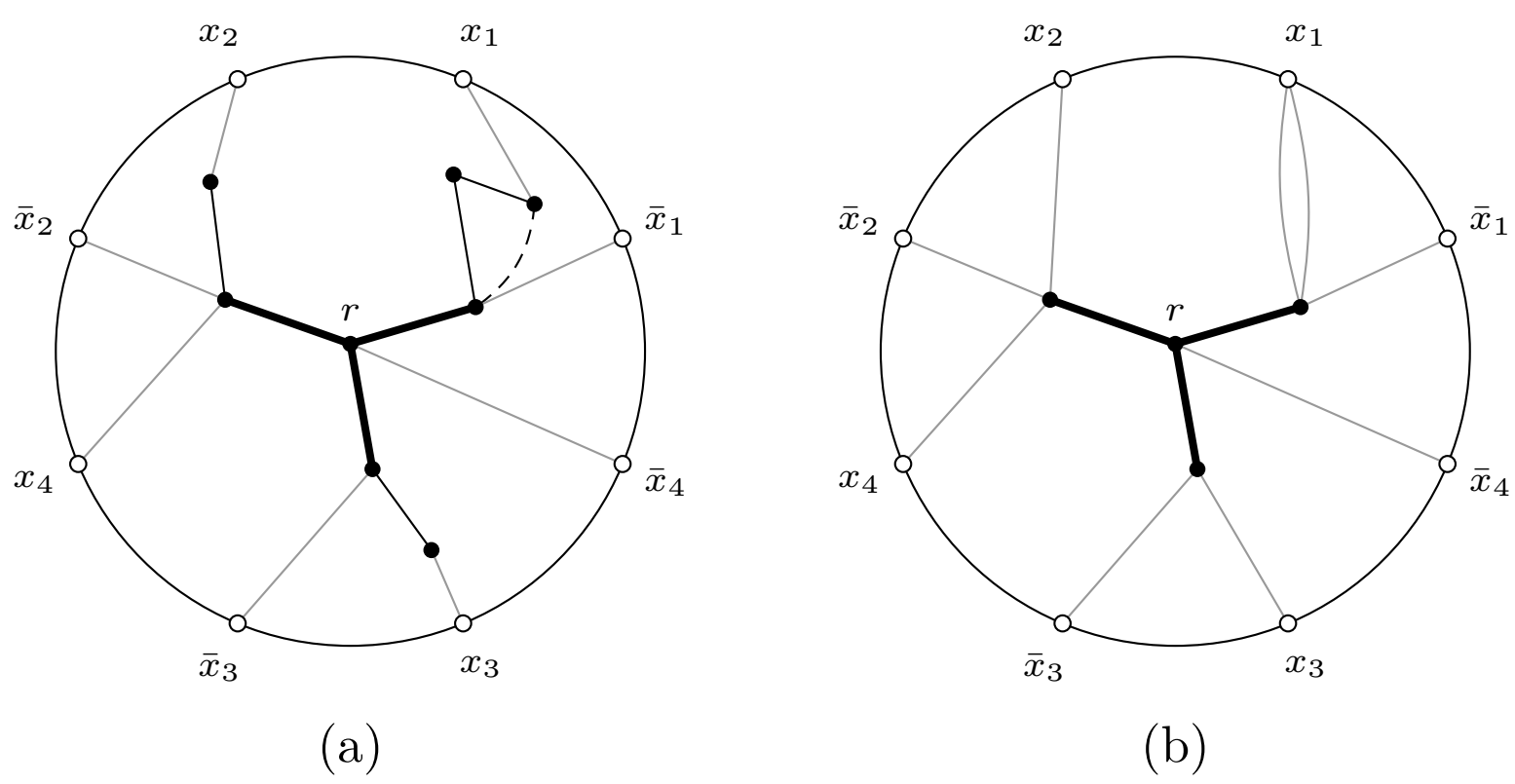
\begin{figure}[t]
\centering
\begin{tikzpicture}[
  vtx/.style={circle,fill,inner sep=1.1pt},
  cvtx/.style={circle,draw,fill=white,inner sep=1.1pt},
  rig/.style={line width=1.5pt},
  tre/.style={black},
  ntr/.style={black,densely dashed},
  half/.style={black!40},
  lbl/.style={font=\scriptsize}]
\newcommand\common{%
  \draw (0,0) circle (2);
  \foreach \k in {0,...,7} \coordinate (c\k) at ({22.5+45*\k}:2);
  \coordinate (r) at (0,0.05);   \coordinate (a) at (-0.85,0.35);
  \coordinate (b) at (0.85,0.3); \coordinate (c) at (0.15,-0.8);
  \draw[rig] (a)--(r)--(b) (r)--(c);
}
\newcommand\finish{%
  \foreach \p in {r,a,b,c} \node[vtx] at (\p) {};
  \foreach \k in {0,...,7} \node[cvtx] at (c\k) {};
  \node[lbl,above=1pt] at (r) {$r$};
  \foreach \k/\t in {1/x_1,0/\bar x_1,2/x_2,3/\bar x_2,6/x_3,5/\bar x_3,4/x_4,7/\bar x_4}
    \node[lbl] at ({22.5+45*\k}:2.32) {$\t$};
}
\begin{scope}
  \common
  \coordinate (z1) at (-0.95,1.15); \coordinate (z2) at (0.7,1.2);
  \coordinate (z3) at (1.25,1.0);   \coordinate (z4) at (0.55,-1.35);
  \draw[tre] (a)--(z1) (b)--(z2)--(z3) (c)--(z4);
  \draw[ntr] (z3) to[bend left=25] (b);
  \draw[half] (z1)--(c2) (z3)--(c1) (z4)--(c6)
              (a)--(c3) (a)--(c4) (b)--(c0) (c)--(c5) (r)--(c7);
  \foreach \p in {z1,z2,z3,z4} \node[vtx] at (\p) {};
  \finish
  \node at (0,-2.7) {(a)};
\end{scope}
\begin{scope}[xshift=5.6cm]
  \common
  \draw[half] (a)--(c2) (a)--(c3) (a)--(c4)
              (b) to[bend left=12] (c1) (b) to[bend right=12] (c1)
              (b)--(c0) (c)--(c6) (c)--(c5) (r)--(c7);
  \finish
  \node at (0,-2.7) {(b)};
\end{scope}
\end{tikzpicture}
\caption{A gadget with a grounded rooted tree, whose rigid tree is thick (a), and its skeleton (b).}
\label{fig:skeleton}
\end{figure}

Proposition~\ref{prop:gadget} is proved as follows, essentially. Call an edge $e$ of $T$ \emph{rigid} if both components of $T-e$ are rich. The rigid edges form a subtree $R$ containing $r$, or $R=\{r\}$. Each component of $T-V(R)$ is not rich, so its halves (if any) go to one vertex of $C$, and, by depth-firstness, its other edges and the matches of its halves end on $R$. So the rest of $H$ splits into pieces attached to $R$, each containing at most one vertex of $C$. Contracting the pieces at each vertex $x$ of $C$ into $x$ gives the \emph{skeleton} (Figure~\ref{fig:skeleton}). As both sides of each rigid edge are rich, all embeddings of the skeleton in the disk bounded by $C$ agree, up to mirror, on $R$, on the corners of $R$ entered by the edges to each $x$, and on their order around $x$. Then $H$ is realizable if and only if the skeleton embeds and, for each pair of matched vertices, the pieces at them, glued along their halves, embed in a disk with these corners on the boundary in this order. If one of these disk problems fails, a theorem of Mohar~\cite[Theorem~3.1]{Mohar} gives a small obstruction to it, and a few paths of $T$ lift it to the desired subgraph of $H$.

Groundedness can be defined directly on rooted spanning trees $(T,r)$ of $G$. Label each oriented edge $e$ of $G$ by the homotopy class $\langle e\rangle$, in $\pi_1(S,f(r))$, of the image under $f$ of the loop that follows $T$ from $r$ to the tail of $e$, then $e$, then $T$ back to $r$. A set $U$ of vertices of $G$ is \emph{rich for $f$} if two oriented edges with tails in $U$ have distinct non-trivial labels, and $(T,r)$ is \emph{grounded for $f$} if it is grounded for this notion of richness.

\begin{lemma}[Appendix~\ref{app:transfer}]\label{lem:transfer}
Let $G$ be a connected graph, $(T,r)$ be a rooted spanning tree of $G$, $f:G\to S$ a map, and $\lambda\circ\mu$ the factorization of $f$ along $T$. If $(T,r)$ is grounded for $f$, then it is grounded for the gadget of this factorization.
\end{lemma}

Such a rooted tree may not exist, for example if $f$ is contractible. More generally, we call $f$ \emph{cyclic} if its image is a cyclic subgroup of the fundamental group of $S$.

\begin{proposition}[Section~\ref{sec:grounding}]\label{prop:grounding}
Let $G$ be a connected graph and $f:G\to S$. If $f$ is not cyclic, and if $f|_O$ can be untangled for every cycle $O$ of $G$, then $G$ has a rooted spanning tree that is depth-first and grounded for $f$.
\end{proposition}

Proposition~\ref{prop:grounding} is proved as follows. In a depth-first rooted tree, the labels of the edges of each block of $G$ outside the tree generate, up to conjugacy, the classes of the closed walks in that block. Every set of vertices containing a block with a non-contractible cycle is rich. If two blocks have non-contractible cycles, rooting the tree at a vertex that separates them, or lies on both, makes it grounded. Otherwise, some block is not cyclic and contains a theta whose paths $X,Y$, completed by the third path $M$, give distinct non-trivial classes. Since every cycle can be untangled, a longest such $M$ lets a depth-first search follow $M$, then $X$, then $Y$. This makes both sides of the first edge of $M$ rich. The cyclic case is handled separately:

\begin{lemma}[Appendix~\ref{app:cyclic}]\label{lem:annular}
Let $G$ be a connected graph and $f : G \to S$. If $f$ is cyclic and cannot be untangled, then $G$ has a subgraph $G'$ with at most $16$ topological vertices such that $f|_{G'}$ cannot be untangled.
\end{lemma}

\begin{proof}[Proof of Theorem~\ref{thm:main}]
It suffices to find $G'\subseteq G$ such that $f|_{G'}$ cannot be untangled, with cycle rank at most $37$ or at most $72$ topological vertices. Indeed, by Section~\ref{sec:prelim}, removing a vertex of degree at most one from $G'$ preserves both its cycle rank and the fact that $f$ cannot be untangled on it. Doing so repeatedly gives a non-empty graph of minimum degree at least two, so with at most $2\cdot37-2=72$ topological vertices if its cycle rank is at most $37$.

Let $\lambda\circ\mu$ be the factorization of $f$ along a spanning forest $T$ of $G$. If $\lambda$ cannot be untangled, Lemma~\ref{lem:loops} gives at most four loops of $L$ on which it cannot be untangled. Let $G'$ consist of one edge sent around each of them and, in each component of $G$, the smallest subtree of $T$ containing their ends. Removing these edges leaves a forest, so $G'$ has cycle rank at most four, and an untangling of $f|_{G'}$ would give one of $\lambda$ on these loops, by contracting $T\cap G'$. So, whatever the choice of $T$, we may assume that $\lambda$ is an embedding, up to homotopy.

If $f$ could be untangled on each component of $G$, then $\mu$ would be a weak embedding on each component, by Lemma~\ref{lem:factorization-untangle}, hence on $G$ itself, as distinct components of $G$ are sent to distinct components of $L$, and $f$ could be untangled. So we may replace $G$ by a component on which $f$ cannot be untangled, and assume that $G$ is connected. If some cycle $O$ of $G$ cannot be untangled, take $G'=O$. If $f$ is cyclic, Lemma~\ref{lem:annular} concludes. Otherwise, by Proposition~\ref{prop:grounding}, we choose for $T$ a rooted spanning tree that is depth-first and grounded for $f$. By Lemma~\ref{lem:transfer}, it is grounded for the gadget $(H,C)$ of the factorization. As its edges are sent to the vertex of $L$, it is also a spanning tree of $H-V(C)$, and it is depth-first in the joined graph of $H$, which is $G$. As $H-V(C)\supseteq T$ is connected, $H$ is not realizable, by Lemma~\ref{lem:gadget}. Proposition~\ref{prop:gadget} gives a non-realizable $H'$ of rank at most $37$ with $H'-V(C)$ connected, and, by Lemma~\ref{lem:gadget}, $f$ cannot be untangled on the corresponding $G'$.
\end{proof}

\section{Preliminaries}\label{sec:prelim}

We use standard notions of graph theory~\cite{Diestel,MoharThomassen} and low-dimensional topology~\cite{Stillwell,FarbMargalit}. The fundamental group of $S$, a compact orientable surface, has no element of order two.

\begin{lemma}[{Epstein~\cite[Theorem~4.2]{Epstein}}]\label{lem:epstein}
A non-contractible simple closed curve on $S$ is not homotopic to $c^k$ for any closed curve $c$ and $k\ge2$.
\end{lemma}

We view graphs as topological spaces that may have loops and parallel edges. The \emph{cycle rank} of a graph is its number of edges, minus its number of vertices, plus its number of connected components. A vertex $v$ is a \emph{cut vertex} of a connected graph if its edges split into two non-empty parts sharing only $v$. A \emph{block} is a maximal connected subgraph without cut vertex. The blocks and cut vertices of a connected graph form a tree, the \emph{block--cut tree}. A \emph{Kuratowski subgraph} is a subdivision of $K_5$ or $K_{3,3}$. A plane embedding of a graph containing a cycle $C$ has \emph{outer face} $C$ if all of the graph lies in the closed disk bounded by $C$. Deleting a vertex of degree at most one preserves the cycle rank, and whether a map can be untangled, as the vertex and its edge can be redrawn near the rest of the graph.

\section{Grounded depth-first trees yield small obstructions}\label{sec:core}

This section proves Proposition~\ref{prop:gadget}. Fix a gadget $(H,C)$, and a rooted spanning tree $(T,r)$ of $H-V(C)$ that is grounded and depth-first in the joined graph of $H$. Assume that $H$ is not realizable. We search for a \emph{witness}: a non-realizable $H'\subseteq H$ of rank at most $37$ such that $H'-V(C)$ is connected. As a preliminary:

\begin{lemma}[Appendix~\ref{app:core}]\label{lem:merge}
Let $F\subseteq H$, and $U$ a subtree of $T$. Joining components of $F-V(C)$ by paths of $U$ relatively disjoint from $F$ can make the vertices of $F$ on $U$ pairwise connected in $F-V(C)$. This adds no half, and leaves the cycle rank of $F-V(C)$ unchanged.
\end{lemma}
 
\subsection{The rigid tree}\label{subsec:rigid-tree}
 
An edge $e$ of $T$ is \emph{rigid} if both components of $T-e$ are rich. Let $R$ be the subgraph of $T$ formed by the rigid edges, or $R=\{r\}$ if there is none. A \emph{hanging subtree} is a component of $T-V(R)$.
 
\begin{lemma}\label{lem:structure}
$R$ is connected and contains $r$. Each hanging subtree $Z$ consists of the descendants of a vertex whose parent $p$ lies in $R$. The halves at $Z$ go to at most one vertex of $C$. Each edge of $H-V(C)$ leaving $Z$, and the match of each half at $Z$, ends on the path of $T$ from $p$ to $r$.
\end{lemma}
 
\begin{proof}
If two edges of $T$ are rigid, so is every edge between them, as each of its sides contains a side of one of them. If some component $K$ of $T-r$ is rich, then so is $V(T)\setminus K$, by groundedness, so the edge from $r$ to $K$ is rigid, and $R$ is a subtree containing $r$. Otherwise, each edge of $T$ has a side contained in a component of $T-r$, which is not rich, so $R=\{r\}$.
 
Let $z$ be the vertex of $Z$ closest to $r$, and $p$ its parent. As $R$ is connected and contains $r$, $p$ lies in $R$, and the descendants of $z$ avoid $R$. So $Z$ consists of the descendants of $z$. It is not rich: if no component of $T-r$ is rich, it lies in one; otherwise, the edge $pz$ is not rigid, and its side containing $r$ is rich, as it contains a side of a rigid edge at $r$. So the halves at $Z$ go to one vertex of $C$.
 
Let $u\in V(Z)$. As $(T,r)$ is depth-first in the joined graph of $H$, the edges of $H-V(C)$ at $u$, and the matches of the halves at $u$, end at ancestors or descendants of $u$. Those outside $Z$ lie on the path from $p$ to $r$. A match of a half at $u$ is not at $Z$, as it goes to a different vertex of $C$.
\end{proof}

For a hanging subtree $Z$, let $p_Z$ be the parent of its vertex closest to $r$, and $c_Z$ the vertex of $C$ reached by its halves, if any.
 
\subsection{Pieces}\label{subsec:pieces}
 
Let $F\subseteq H$, and let $R'\subseteq F$ be a subtree of $R$. An \emph{$R'$-piece} of $F$ is an edge of $F-E(R')$ with both ends in $R'\cup C$, or a component of $F-V(R')-V(C)$ together with its edges to $R'\cup C$. The pair $(F,R')$ is \emph{admissible} if $F-V(C)$ is connected, every $R'$-piece contains at most one vertex of $C$, every $R'$-piece avoiding $C$ meets $R'$ at least twice, and, for each leaf $q$ of $R'$, two $R'$-pieces containing $q$ contain distinct vertices of $C$.
 
\begin{lemma}\label{lem:cleaning}
Assume that $H-V(C)$ is planar. Deleting the $R$-pieces of $H$ that avoid $C$ and meet $R$ only once preserves realizability, and makes $(H,R)$ admissible.
\end{lemma}
 
\begin{proof}
These pieces are planar, have no halves, and each meets the rest of $H$ at one vertex. So, in a realization of the rest, they can be drawn in small disks around these vertices, with these vertices on their outer faces, and deleting them preserves realizability. As these pieces have no halves, no rich set changes, and $T$ minus the deleted vertices is a spanning tree of the new $H-V(C)$, still grounded and depth-first in its joined graph. So Lemma~\ref{lem:structure} still holds, with the same $R$.
 
The new pair $(H,R)$ is admissible. Indeed, $H-V(C)$ contains $T$, so it is connected. Each piece contains at most one vertex of $C$, by the description above, and the pieces avoiding $C$ now meet $R$ twice. Let $q$ be a leaf of $R$. If $R=\{q\}$, then $H$ has a half, as $V(T)$ is rich; its match goes to another vertex of $C$, and every piece contains $q$. Otherwise, let $e$ be the edge of $R$ at $q$. As $e$ is rigid, the component $U$ of $T-e$ containing $q$ is rich. As $q$ is a leaf of $R$, the other vertices of $U$ lie in hanging subtrees $Z$ with $p_Z=q$. So the halves at $U$ lie in pieces containing $q$, and they reach two vertices of $C$.
\end{proof}
 
\subsection{Disk obstructions}\label{subsec:disk}
 
\begin{lemma}\label{lem:disk}
If a connected graph $F$ has no plane embedding with a given cycle $C$ as outer face, then neither does $C\cup\Omega$, for some $\Omega\subseteq F-E(C)$ of cycle rank at most $6$, with at most $4$ vertices on $C$ and at most $8$ edges at them, such that $\Omega-V(C)$ has at most $2$ components.
\end{lemma}

\begin{proof}
Subdividing $C$, we may assume $|C|\ge3$. Let $F^+$ be $F$ plus a vertex $v$ adjacent to all of $C$. By~\cite[Theorem~3.1]{Mohar}, some $\Omega\subseteq F-E(C)$ such that $C\cup\Omega$ has no plane embedding with outer face $C$ is of one of the following types: (a) two disjoint paths, internally disjoint from $C$, whose ends alternate along $C$; (b) a subdivided $K_{2,3}$, with three disjoint paths, internally disjoint from $C$ and possibly trivial, from its three vertices of degree two to $C$; (c) a Kuratowski subgraph of a $3$-connected component of $F^+$ not containing $C+v$. 

Types (a) and (b) are as claimed. In type (a), $\Omega$ has cycle rank zero and four edges at $C$. In type (b), it has cycle rank two and at most six edges at $C$, and $\Omega-V(C)$ contains the two vertices of degree three of the $K_{2,3}$, and is connected unless its three other vertices lie on $C$. 

In type (c), $F^+$ has a separation $(F_1,F_2)$, with $\Sigma$$=V(F_1\cap F_2)$ of size at most two, $F_2$ connected and containing $C+v$, such that $F_1$ plus an edge joining the vertices of $\Sigma$, if there are two, is not planar: take $\Sigma$ separating the two $3$-connected components in question, the first being a topological minor of $F_1$ plus that edge. Then $F_1$ meets $C+v$ only in $\Sigma$. If $v\in\Sigma$, then $v$ has at most one neighbor in $F_1$, so $F_1-v$ is not planar, and we take for $\Omega$ a Kuratowski subgraph of it. Otherwise, let $\mathcal K$ be a Kuratowski subgraph of $F_1$, plus the edge $st$ if $\Sigma=\{s,t\}$. If $\mathcal K$ uses $st$, replace $st$ by a path of $F_2$ from $s$ to $t$, and the part of this path between its first and last vertices $a,b$ on $C$, if any, by an arc of $C$; this avoids $v$, whose neighbors lie on $C$. Then $C\cup\mathcal K$ is not planar, and $\Omega=\mathcal K-E(C)$, minus its isolated vertices, meets $C$ only in $s,t,a,b$.

In type (c), $\Omega$ is a Kuratowski subgraph, of maximum degree four, possibly minus an edge $st$, with two paths attached by one end, so it has cycle rank at most six. It has at most eight edges at $C$: at most four at each of $s$ and $t$, except that if $a\ne s$, then $s\notin C$ and $a$ has degree one, and likewise for $b$ and $t$. Finally, $\Omega-V(C)$ is a subdivided $3$-connected graph minus at most two vertices or one edge and some vertices of degree one, with paths attached by one end, so it has at most two components.
\end{proof}
 
\subsection{The skeleton}\label{subsec:skeleton}
 
Let $(F,R')$ be admissible. For $x\in V(C)$, let $F_x$ be the union of the $R'$-pieces of $F$ containing $x$; then $F_x-V(R')$ is connected and contains $x$. The \emph{skeleton} $K(F,R')$ is obtained from $F\cup C$ by contracting $F_x-V(R')$ into $x$, for each $x\in V(C)$. It consists of $C$, of the \emph{core} $K_0$ formed by $R'$ and the $R'$-pieces avoiding $C$, and of \emph{star edges}: for each edge of $F_x$ at a vertex $a$ of $R'$, a star edge from $a$ to $x$. An \emph{embedding} of the skeleton is a plane embedding with outer face $C$.
 
\begin{lemma}\label{lem:skeleton}
If $(H,R)$ is admissible and $K(H,R)$ has no embedding, then $H$ has a witness.
\end{lemma}
 
\begin{proof}
The skeleton is connected, as $H-V(C)$ is. So Lemma~\ref{lem:disk} gives $\Omega\subseteq K(H,R)-E(C)$ of cycle rank at most six such that $C\cup\Omega$ has no embedding; its edges at $C$ are star edges, and there are at most eight of them. Each star edge of $\Omega$, from $a$ to $x$, comes from a half $ax$, or from an edge $au$ with $u$ in a hanging subtree $Z$ with $c_Z=x$. Let $J$ be the union of the following: 1) $\Omega$ minus its star edges and its vertices on $C$; 2) for each star edge of $\Omega$, such a half $ax$ or edge $au$; 3) a half at a vertex of each such $Z$; 4) the matches of these halves. Each hanging subtree $Z$ containing halves and met by $J$ contains a half of $J$. Apply Lemma~\ref{lem:merge} to the part of $J$ in $Z$, with $U=T[Z]$; as this part has no edge, it becomes a subtree of $T$.

Assume that $J$ is realizable. In a realization, contract into $c_Z$ the part of $J$ in each hanging subtree $Z$ containing halves and met by $J$, with its halves; this part is connected and has a half at $c_Z$. This turns each edge $au$ into a star edge, and embeds a graph containing $C\cup\Omega$, a contradiction. So $J$ is not realizable, and stays so when Lemma~\ref{lem:merge}, with $U=T$, makes $J-V(C)$ connected.

The result is a witness. Indeed, $J$ has at most eight pairs of matched halves, at most one for each star edge of $\Omega$, as the halves of item~3 replace the edges $au$. Before the last application of Lemma~\ref{lem:merge}, contract the subtrees of $J$ in hanging subtrees, and identify those in hanging subtrees $Z$ with the same vertex $c_Z$ with that vertex. This maps $J-V(C)$ to a subgraph of $\Omega$, plus isolated vertices. Contracting trees preserves the cycle rank, and identifying vertices does not decrease it. So $J-V(C)$ has cycle rank at most six, also after Lemma~\ref{lem:merge}, and $J$ has rank at most $6+8=14$.
\end{proof}
 
\subsection{Local graphs}\label{subsec:local}
 
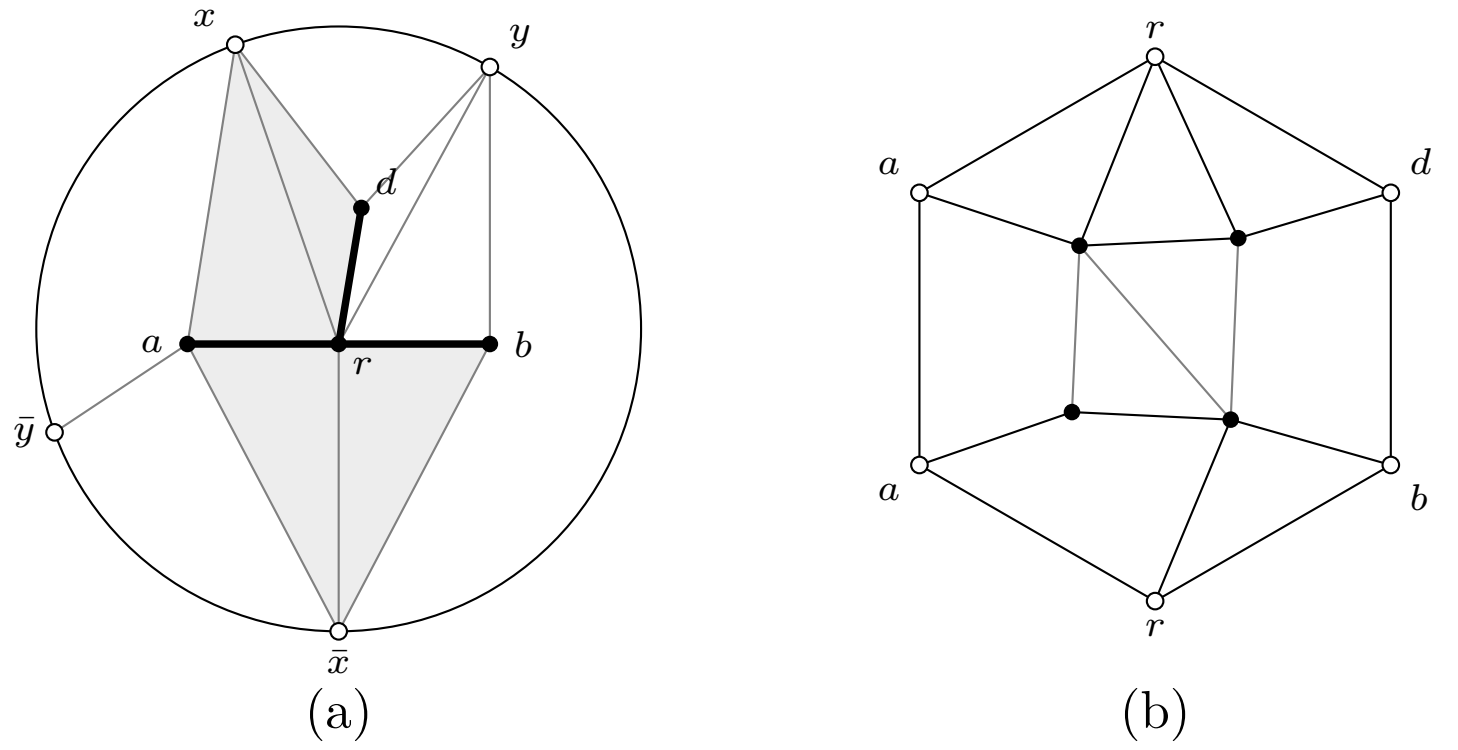
\begin{figure}[t]
\centering
\begin{tikzpicture}[
  vtx/.style={circle,fill,inner sep=1.1pt},
  cvtx/.style={circle,draw,fill=white,inner sep=1.1pt},
  rig/.style={line width=1.4pt},
  half/.style={gray},
  lbl/.style={font=\scriptsize}]
\begin{scope}
  \coordinate (x) at (110:2); \coordinate (y) at (60:2);
  \coordinate (xb) at (270:2); \coordinate (yb) at (200:2);
  \coordinate (r) at (0,-0.1); \coordinate (a) at (-1,-0.1);
  \coordinate (b) at (1,-0.1); \coordinate (d) at (0.15,0.8);
  \fill[gray!14] (x)--(a)--(r)--(d)--cycle;
  \fill[gray!14] (xb)--(a)--(r)--(b)--cycle;
  \draw (0,0) circle (2);
  \draw[half] (x)--(a) (x)--(r) (x)--(d) (y)--(d) (y)--(r) (y)--(b)
              (xb)--(a) (xb)--(r) (xb)--(b) (yb)--(a);
  \draw[rig] (a)--(r)--(b) (r)--(d);
  \foreach \p in {a,r,b,d} \node[vtx] at (\p) {};
  \foreach \p in {x,y,xb,yb} \node[cvtx] at (\p) {};
  \node[lbl,above left] at (x) {$x$}; \node[lbl,above right] at (y) {$y$};
  \node[lbl,below] at (xb) {$\bar x$}; \node[lbl,left] at (yb) {$\bar y$};
  \node[lbl,left=1pt] at (a) {$a$}; \node[lbl,below right=-1pt] at (r) {$r$};
  \node[lbl,right=1pt] at (b) {$b$}; \node[lbl,above right=-1pt] at (d) {$d$};
  \node at (0,-2.55) {(a)};
\end{scope}
\begin{scope}[xshift=5.4cm]
  \foreach \n/\ang in {ca/150, cr/90, cd/30, cb/330, crb/270, cab/210}
    \coordinate (\n) at (\ang:1.8);
  \draw (ca)--(cr)--(cd)--(cb)--(crb)--(cab)--cycle;
  \coordinate (p1) at (-0.5,0.55); \coordinate (p2) at (0.55,0.6);
  \coordinate (q1) at (0.5,-0.6); \coordinate (q2) at (-0.55,-0.55);
  \draw (p1)--(ca) (p1)--(cr) (p2)--(cr) (p2)--(cd) (p1)--(p2);
  \draw (q1)--(cb) (q1)--(crb) (q2)--(cab) (q1)--(q2);
  \draw[half] (p1)--(q2) (p1)--(q1) (p2)--(q1);
  \foreach \p in {p1,p2,q1,q2} \node[vtx] at (\p) {};
  \foreach \p in {ca,cr,cd,cb,crb,cab} \node[cvtx] at (\p) {};
  \node[lbl,above left] at (ca) {$a$}; \node[lbl,above] at (cr) {$r$}; \node[lbl,above right] at (cd) {$d$};
  \node[lbl,below right] at (cb) {$b$}; \node[lbl,below] at (crb) {$r$}; \node[lbl,below left] at (cab) {$a$};
  \node at (0,-2.55) {(b)};
\end{scope}
\end{tikzpicture}
\caption{The fans of $x$ and $\bar x$ in a skeleton (a), and the local graph at $\{x,\bar x\}$ (b).}
\label{fig:local}
\end{figure}

Let $(F,R')$ be admissible, and $\psi$ an embedding of its skeleton. Around a vertex $a$ of $R'$, the edges of $R'$ at $a$ cut a small disk centered at $a$ into \emph{sectors}; there is a single sector if $a$ is a leaf of $R'$. A \emph{corner} is a vertex $a$ of $R'$ with a sector at $a$. A star edge with end $a$ in $R'$ \emph{enters} the corner of the sector containing its initial part at $a$. See Figure~\ref{fig:local}.
 
\begin{lemma}\label{lem:fans}
Let $x\in V(C)$ have star edges $s_1,\dots,s_k$, in clockwise order. The faces of $\psi$ between $s_i$ and $s_{i+1}$ at $x$, with $s_2,\dots,s_{k-1}$, form an open disk, the \emph{fan} of $x$ (a thin neighborhood of $s_1$ if $k=1$), bounded by $s_1$, $s_k$, and a walk of $K_0$ visiting the corners entered by $s_1,\dots,s_k$ once each, in order.
\end{lemma}
 
\begin{proof}
The face between $s_i$ and $s_{i+1}$ at $x$ meets $C$ only at this angle: otherwise, a curve inside it from this angle to another point or angle of $C$ would separate the ends of $s_i$ and $s_{i+1}$ in $R'$. So the fan is an open disk bounded by $s_1$, $s_k$, and a walk $W$ of $K_0$.
 
Let $s,s'$ be star edges entering the same corner at $a$, consecutive around $a$ among such edges. With an arc of $C$, they bound a disk $\Delta$ containing the angle between them at $a$, and thus no edge of $R'$ at $a$. As $R'$ is connected, $\Delta$ meets $R'$ only at $a$. So $\Delta$ contains no piece avoiding $C$, as these meet $R'$ twice, and no star edge: $\Delta$ is a face of $\psi$. Hence the star edges entering a corner of $R'$ end at the same corner of the outer face of $K_0$.

Finally, $W$ is part of the boundary walk of the outer face of $K_0$, which visits each of its corners once, and $W$ visits no corner twice: otherwise, it would go all around $K_0$, and all leaves of $R'$ but at most one would have star edges only to $x$, contradicting admissibility.
\end{proof}
 
Let $X=\{x,\bar x\}$ be a pair of matched vertices of $C$. The \emph{local graph} $G_X$ (Figure~\ref{fig:local}) is obtained from $F_x\cup F_{\bar x}$ as follows. Join each pair of matched halves into a single edge, and delete $x$ and $\bar x$. Then detach the edges at the vertices of $R'$, and reattach those of $F_x$ whose star edges enter the same corner to a common new vertex, a \emph{corner of $x$}; likewise for $\bar x$. The \emph{boundary} $C_X$ is the cycle through the corners of $x$, in the clockwise order of their star edges around $x$, then through the corners of $\bar x$, likewise. By Lemma~\ref{lem:fans}, this order is well defined.
 
\begin{proposition}\label{prop:local}
$F$ is realizable if and only if its skeleton has an embedding $\psi$ for which, for each matched pair $X$, $G_X\cup C_X$ has a plane embedding with outer face $C_X$.
\end{proposition}
 
\begin{proof}
Given $\psi$ and these embeddings, draw $C$ and $K_0$ as in $\psi$. Fix $X=\{x,\bar x\}$. In $G_X$, the part coming from $F_x$, with the corners of $x$, is connected, as each piece meets $R'$, and contains an arc of $C_X$; likewise for $\bar x$. So the joined halves and two edges of $C_X$ form a minimal cut, and some curve in the disk crosses exactly these edges, once each. Cutting along this curve and contracting each side into a point gives embeddings of $F_x$ and $F_{\bar x}$ in disks, with $x$, resp.\ $\bar x$, and their corners on the boundary, and with reversed orders of the halves at $x$ and $\bar x$. Up to mirror, the corners are in the order of Lemma~\ref{lem:fans}. So we can map these disks into the closures of the fans of $x$ and $\bar x$, sending each corner to the point where the walk of the fan visits it. Doing so for every $X$ gives a realization of $F$.
 
Conversely, consider a realization of $F$. For each $x\in V(C)$, the graph $F_x-V(R')$ meets $C$ only at $x$, so the rest of the drawing lies in its face containing $C-x$. Hence contracting these graphs gives an embedding $\psi$ of the skeleton, and $F_x$ is drawn in the closure of the fan of $x$, each edge at $R'$ at the corner entered by its star edge. Joining the fans of $x$ and $\bar x$ by a band at $x$ and $\bar x$ embeds $G_X\cup C_X$ with outer face $C_X$: the joined halves run along the band without crossing, as their orders at $x$ and $\bar x$ are reversed.
\end{proof}
 
\subsection{Rigidity}\label{subsec:rigidity}
 
\begin{lemma}\label{lem:rigidity}
Let $(F,R')$ be admissible, and $\psi,\psi'$ two embeddings of its skeleton. For each matched pair $X$, the local graphs of $\psi$ and $\psi'$ at $X$ coincide, up to reversing their boundary.
\end{lemma}
 
\begin{proof}
Let $C^*$ be the set of vertices of $C$ with star edges. Let $Q$ be the graph formed by $R'$, the star edges, a cycle $C'$ through $C^*$ in the cyclic order of $C$, and a vertex $v$ adjacent to all of $C^*$. Deleting $K_0-R'$ from $\psi$, suppressing the vertices of $C$ outside $C^*$, and adding $v$ outside $C$ gives an embedding of $Q$. Its rotation system determines the local graphs: the corner entered by a star edge is read from the rotation at its end in $R'$, and the order of the star edges around $x$ from the rotation at $x$. Reversing all rotations reverses all boundaries. It suffices that $Q$ has a unique embedding, up to mirror and to the order of parallel star edges.
 
Merge parallel edges and suppress vertices of degree two. By Whitney's theorem~\cite{Whitney}, it suffices that the result is $3$-connected, or has at most three vertices. So let $p,q\in V(Q)$; we show that the vertices of degree at least three of $Q-\{p,q\}$ lie in one component. First, $(C'+v)-\{p,q\}$ is connected, unless $C^*=\{p,q\}$, in which case $v$ has degree two. Second, let $w$ be a vertex of $R'-\{p,q\}$ of degree at least three. Its component in $R'-\{p,q\}$ contains a leaf of $R'$, which has star edges to two vertices of $C^*$, or is part of the path of $R'$ between $p$ and $q$, on which $w$ has degree two in $R'$, and thus a star edge. In both cases, $w$ is joined to $C^*-\{p,q\}$ in $Q-\{p,q\}$, unless $p,q\in C^*$; then $R'$ is connected, with star edges to all of $C^*$.
 
Finally, parallel star edges are consecutive around their ends: otherwise, their $2$-cycle would separate vertices of degree at least three, as no path of $Q$ through vertices of degree two joins a vertex of $R'$ to one of $C^*$. So they enter the same corner.
\end{proof}

\subsection{Lifting}\label{subsec:lifting}

We now turn an obstruction in a local graph into a witness, by completing subgraphs of $H$ into admissible pairs of bounded rank. A set $N\subseteq V(R)$ \emph{anchors} a subgraph $F\subseteq H$ if it contains $r$, the vertices of $F$ in $R$, and the vertex $p_Z$ for each hanging subtree $Z$ meeting $F$.

\begin{lemma}\label{lem:closure}
Let $F\subseteq H$ meet $R$, be anchored by $N$, and have no edge with both ends in $R$, and let $m$ be the number of components of $F-V(C)-V(R)$. There is $H'\supseteq F$ whose rank exceeds that of $F$ by at most $|V(F)\cap V(R)|+2m+2|N|+5$, such that $(H',R')$ is admissible for $R'=H'\cap R\supseteq N$, and every $R'$-piece of $H'$ meeting a hanging subtree $Z$ contains $c_Z$.
\end{lemma}

\begin{proof}
Let $q$ be a leaf of $R$. As $(H,R)$ is admissible, two $R$-pieces containing $q$ contain distinct vertices of $C$. A \emph{link} at $q$ is an edge at $q$ of each of them. If $q\ne r$, then $q$ has no descendant in $R$. So the links at $q$, and the matches of those that are halves, end at ancestors of $q$, or in hanging subtrees $Z$ with $p_Z=q$.

Let $N'$ be $N$ plus, if $r$ is a leaf of $R$, at most two vertices, as follows. Each link at $r$ has, besides $r$, one end in $T$: its other end, or that of its match if it is a half. Add this end if it lies in $R$, or $p_Z$ if it lies in a hanging subtree $Z$. Let $R_0$ be the smallest subtree of $R$ containing $N'$. Each component of $R-E(R_0)$ meets $R_0$ in one vertex. If $R_0$ has at least two vertices, for each leaf $q$ of $R_0$ that is not a leaf of $R$, add to $R_0$ a path from $q$ to a leaf of $R$ in the component of $R-E(R_0)$ at $q$, which has another vertex. If $R_0$ is a single vertex, not a leaf of $R$, it is $r$; add two paths from it to leaves of $R$, starting with distinct edges. The resulting tree $R'$ contains $r$, and its $\ell\le|N|+2$ leaves are leaves of $R$.

Let $H'$ consist of $F$, $R'$, the links at the leaves of $R'$, a half at a vertex of each hanging subtree met by $F$ or by these links, and the matches of all these halves. For each hanging subtree $Z$ met by $H'$, apply Lemma~\ref{lem:merge} to the part of $H'$ in $Z$, plus $p_Z$, with $U$ the subtree of $T$ formed by $Z$ and $p_Z$. By Lemma~\ref{lem:structure} and the above, all ends in $R$ of the added edges, and all these vertices $p_Z$, lie in $R'$; so $H'\cap R=R'$.

The $R'$-pieces of $H'$ are its halves at $R'$ and, for each hanging subtree $Z$ met by $H'$, its part in $Z$, which is connected, joined to $p_Z$, and contains a half to $c_Z$. So $H'-V(C)$ is connected, and each piece contains exactly one vertex of $C$. At each leaf of $R'$, the links lie in two pieces containing distinct vertices of $C$.

Finally, we bound the rank of $H'$, the cycle rank of the subgraph of $G$ corresponding to $H'$. Let $\rho$ be the rank of $F$, and $s=|V(F)\cap V(R)|$. Adding $R'$ to $F$ increases the rank by at most $s-1$, as $R'$ is a tree meeting $F$ in $s$ vertices and no edge. We then add edges, each link that is not a half, and each pair of matched halves, counting as one edge, and then the paths of Lemma~\ref{lem:merge}. The increase of the cycle rank is at most the number of these edges with no new end, plus the number of these paths. In a hanging subtree $Z$, Lemma~\ref{lem:merge} adds at most one path for each component of $F-V(C)-V(R)$ and for each new vertex not adjacent to $p_Z$. So each component of $F-V(C)-V(R)$ costs at most two: one path, and one for the half at its hanging subtree, as an edge or through the path of its new end. A link at a leaf $q\ne r$ costs at most one: an edge, or a path for the new end of its match in a hanging subtree $Z$; if it is an edge to a new vertex $u$ of $Z$, then $u$ is adjacent to $p_Z=q$, and the link costs nothing but the half at $Z$, which costs at most one. Similarly, a link at $r$ costs at most two. So the rank of $H'$ is at most $\rho+(s-1)+2m+2\ell+2\le\rho+s+2m+2|N|+5$.
\end{proof}

Now fix an admissible pair $(H,R)$ with $H-V(C)$ planar, an embedding $\psi$ of its skeleton, a matched pair $X$, and $\Omega\subseteq G_X-E(C_X)$ as given by Lemma~\ref{lem:disk}. An admissible pair $(H',R')$ with $H'\subseteq H$ \emph{captures} $\Omega$ if its skeleton is contained in $K(H,R)$ and, for the restriction of $\psi$, its local graph at $X$ contains $\Omega$, with the same cyclic order of $\Omega\cap C_X$ on the boundary.

\begin{lemma}\label{lem:lifting}
Some admissible pair $(H',R')$, with $H'$ of rank at most $37$, captures $\Omega$.
\end{lemma}

\begin{proof}
By Lemma~\ref{lem:structure}, the matches of the halves at hanging subtrees end in $R$, so every joined edge has a corner as an end, and $G_X-V(C_X)$ is the disjoint union of the graphs $H[Z]$ for the hanging subtrees $Z$ with $c_Z\in X$. Replace each joined edge of $\Omega$ by its two matched halves, and each corner by its vertex. This gives $F\subseteq H$. Its vertices in $R$ form the set $A$ of vertices of the corners of $\Omega$, it has no edge joining two vertices of $R$, and its other vertices off $C$ form $\Omega-V(C_X)$, whose at most two components each lie in a hanging subtree $Z$ with $c_Z\in X$. And $A\neq\emptyset$, as otherwise $\Omega\subseteq H-V(C)$ would not be planar. The subgraph of $G$ corresponding to $F$ is obtained from $\Omega$ by identifying the corners with the same vertex, so $F$ has rank at most $6+c-|A|$, where $c\le4$ is the number of corners of $\Omega$.

Let $N$ consist of $r$, $A$, the vertices $p_Z$ of these at most two hanging subtrees, and, for each $a\in A$ with $k\ge2$ corners of $\Omega$, $k$ leaves of $R$, one beyond an edge of $R$ at $a$ between each two of these corners consecutive around $a$. As $c\le4$, the vertices of $A$ and these leaves number at most six, so $|N|\le9$. Then $N$ anchors $F$, and Lemma~\ref{lem:closure} gives $(H',R')$, with $H'$ of rank at most $(6+c-|A|)+|A|+4+18+5\le37$.

As $H'\cap R=R'$, each $R'$-piece of $H'$ lies in an $R$-piece of $H$, with the same edges at $R'$. So the skeleton of $(H',R')$ is contained in $K(H,R)$. In the restriction of $\psi$, a corner of $R'$ at $a$ is a union of corners of $R$ at $a$, and the star edges keep their clockwise order. Distinct corners of $\Omega$ at $a$ remain distinct, as $R'$ contains an edge at $a$ between any two of them. So the corners of $\Omega$ are corners of the local graph of $(H',R')$ at $X$, in the same cyclic order on its boundary. By Lemma~\ref{lem:closure}, the vertices of $\Omega$ off $C_X$ lie in pieces containing $x$ or $\bar x$. So this local graph contains $\Omega$.
\end{proof}
 
\subsection{Proof of Proposition~\ref{prop:gadget}}
 
\begin{proof}[Proof of Proposition~\ref{prop:gadget}]
If $H-V(C)$ is not planar, a Kuratowski subgraph of it is a witness, of rank at most six. Otherwise, by Lemma~\ref{lem:cleaning}, we may assume that $(H,R)$ is admissible. If $K(H,R)$ has no embedding, Lemma~\ref{lem:skeleton} gives a witness. Otherwise, let $\psi$ be one. As $H$ is not realizable, Proposition~\ref{prop:local} gives a matched pair $X$ such that $G_X\cup C_X$ has no plane embedding with outer face $C_X$. This graph is connected, as each piece meets $R$. So Lemma~\ref{lem:disk} gives $\Omega$, and Lemma~\ref{lem:lifting} gives an admissible pair $(H',R')$ capturing $\Omega$, with $H'$ of rank at most $37$.
 
Assume that $H'$ is realizable. By Proposition~\ref{prop:local}, some embedding of the skeleton of $(H',R')$ has local graphs with plane embeddings with outer faces their boundaries. By Lemma~\ref{lem:rigidity}, so does the restriction of $\psi$. But its local graph at $X$ contains $\Omega$, with the same cyclic order on the boundary, so it has none. Hence $H'$ is not realizable. As $(H',R')$ is admissible, $H'-V(C)$ is connected, so $H'$ is a witness.
\end{proof}

\section{Grounded depth-first trees exist}\label{sec:grounding}

This section proves Proposition~\ref{prop:grounding}. We fix a connected graph $G$ and a map $f:G\to S$ that is not cyclic, such that $f|_O$ can be untangled for every cycle $O$ of $G$. For a closed walk $W$ in $G$, we write $[W]$ for the free homotopy class of $f\circ W$. Rooted spanning trees are depth-first in $G$, and richness and groundedness are for $f$.

\subsection{Images of blocks}\label{subsec:images}
 
The \emph{image} of a connected subgraph $B$ of $G$ is the subgroup of the fundamental group of $S$ to which the fundamental group of $B$ is mapped by $f$, for an arbitrary basepoint.
 
\begin{lemma}\label{lem:dfs}
Let $(T,r)$ be depth-first. The components of $T-r$ are the vertex sets of the components of $G-r$. For each block $B$ of $G$, $T\cap B$ is a spanning tree of $B$, and the labels of the edges of $B$ not in $T$ generate a conjugate of the image of $B$.
\end{lemma}
 
\begin{proof}
Every edge of $G$ joins a vertex to one of its ancestors, so no edge joins two components of $T-r$; this gives the first statement. Let $uu'$ be an edge of $B$ not in $T$, with $u$ an ancestor of $u'$. With the path of $T$ from $u$ to $u'$, it forms a cycle, which lies in $B$. So, as $B$ is connected, $T\cap B$ is a spanning tree of $B$. Let $t$ be its vertex closest to $r$. For every vertex $u$ of $B$, the path of $T$ from $r$ to $u$ goes through $t$, and then stays in $B$. So the label of an edge of $B$ not in $T$ is the class of its fundamental cycle in $T\cap B$, based at $t$, conjugated by the class of the path of $T$ from $r$ to $t$. These fundamental cycles generate the image of $B$.
\end{proof}
 
As the labels of all the edges generate the image of $G$, which is not cyclic, Lemma~\ref{lem:dfs} shows that some block of $G$ has non-cyclic image, or that two blocks have non-trivial images. We treat these cases separately, using the following consequence of Lemma~\ref{lem:dfs}.
 
\begin{lemma}\label{lem:block-rich}
Let $(T,r)$ be depth-first, and $B$ a block with non-trivial image. Every set of vertices containing $V(B)$ is rich.
\end{lemma}
 
\begin{proof}
By Lemma~\ref{lem:dfs}, some edge of $B$ has a non-trivial label $x$. Oriented both ways, it emanates from the set, with labels $x$ and $x^{-1}$, which are distinct as $\pi_1(S)$ is torsion-free.
\end{proof}
 
\subsection{Two blocks with non-trivial images}\label{subsec:two-blocks}
 
\begin{lemma}\label{lem:two-blocks}
If two blocks of $G$ have non-trivial images, then $G$ has a rooted spanning tree that is both depth-first and grounded for $f$.
\end{lemma}
 
\begin{proof}
Assume first that two blocks $B_1,B_2$ with non-trivial images are disjoint. Let $r$ be the cut vertex following $B_1$ on the path from $B_1$ to $B_2$ in the block--cut tree. Then $r\in V(B_1)$, and $B_2$ lies in a component $K$ of $G-r$ that avoids $B_1$. Let $(T,r)$ be any depth-first rooted spanning tree, and $e$ the edge of $T$ from $r$ to $K$. By Lemma~\ref{lem:dfs}, the two sides of $e$ are $V(K)\supseteq V(B_2)$ and its complement, which contains $V(B_1)$. Both are rich by Lemma~\ref{lem:block-rich}. So $V(G)$ is rich, and, for every rich component $K'$ of $T-r$, $V(G)\setminus K'$ is rich: it is $V(G)\setminus K$ or contains $K$. Hence $T$ is grounded for $f$.
 
Otherwise, the blocks with non-trivial images pairwise intersect, so they share a vertex $v$, by the Helly property of subtrees of the block--cut tree. Let $(T,v)$ be any depth-first rooted spanning tree. Distinct blocks containing $v$ meet distinct components of $G-v$. So every component $K$ of $T-v$ avoids some block with non-trivial image, and $V(G)\setminus K$ is rich, by Lemma~\ref{lem:block-rich}. As $V(G)$ is rich, $T$ is grounded for $f$.
\end{proof}
 
\subsection{A block with non-cyclic image}\label{subsec:tripaths}
 
A \emph{tripath} consists of two vertices $m\ne n$ and three internally disjoint paths: $M$ from $m$ to $n$, and $X$ and $Y$ from $n$ to $m$, such that $x=[MX]$ and $y=[MY]$ are non-trivial and distinct, where, in this subsection, $[MX]$, $[MY]$, and the classes of other closed walks starting with $M$ are taken in $\pi_1(S,f(m))$. Its \emph{length} is the number of edges of $M$.
 
\begin{lemma}\label{lem:no-inverse}
In every tripath, $x\ne y^{-1}$.
\end{lemma}
 
\begin{proof}
If $x=y^{-1}$, then $X$ is homotopic to $M^{-1}Y^{-1}M^{-1}$ relative to its ends. So the cycle $X\cup Y$, read as the closed walk $XY^{-1}$ from $n$, has class $[(M^{-1}Y^{-1})^2]$, the square of a conjugate of $y^{-1}\ne1$. By Lemma~\ref{lem:epstein}, this cycle cannot be untangled, a contradiction.
\end{proof}
 
\begin{lemma}\label{lem:tripath}
If some block of $G$ has non-cyclic image, then $G$ has a tripath.
\end{lemma}
 
\begin{proof}
Any such block $B$ is $2$-connected, as single edges and loops have cyclic images. Take an ear decomposition $B_0\subset\dots\subset B_k=B$, where $B_0$ is a cycle, and $B_i$ is $B_{i-1}$ plus a path $Q$ with distinct ends in $B_{i-1}$ and inner vertices outside it. The image of $B_i$ is generated by that of $B_{i-1}$ and the class of any closed walk going once through $Q$. Choose $i$ minimal such that $B_i$ has non-cyclic image. Then $B_{i-1}$ has image $\langle c\rangle$, with $c\ne1$, as otherwise $B_i$ would have cyclic image, and every closed walk of $B_i$ going once through $Q$, joined to the base point inside $B_{i-1}$, has a class outside $\langle c\rangle$.
 
Let $Z$ be a cycle of $B_{i-1}$ with $[Z]\ne1$, which exists as the classes of cycles generate the image of $B_{i-1}$ up to conjugacy. As $B_{i-1}$ is $2$-connected, the ends of $Q$ are joined to $Z$ by disjoint, possibly trivial, paths, meeting $Z$ only at distinct ends $m$ and $n$. Let $M$ be the path from $m$ to $n$ formed by these paths and $Q$, and let $X$ and $Y$ be the two paths of $Z$ from $n$ to $m$. Then $[MX]$ and $[MY]$ are non-trivial, as $MX$ and $MY$ go once through $Q$. They are distinct, as $[MX]^{-1}[MY]=[X^{-1}Y]=[Z]^{\pm1}$, up to conjugacy, is non-trivial.
\end{proof}
 
\begin{lemma}\label{lem:separated}
In a tripath of maximum length, no path joins an inner vertex of $X$ to an inner vertex of $Y$ while avoiding the rest of $M\cup X\cup Y$.
\end{lemma}
 
\begin{proof}
Assume that such a path $P$ joins $p\in X$ to $q\in Y$. Split $X=X_1X_2$ at $p$ and $Y=Y_1Y_2$ at $q$, and let $e=[MX_1PY_2]$ and $e'=[MY_1P^{-1}X_2]$. Consider the four thetas whose middle paths are $MX_1$, $MY_1$, $X_2M$ and $Y_2M$; they are longer than $M$. Up to conjugacy by a common element within each pair, their pairs of classes are $(x,e)$, $(y,e')$, $(x,e')$ and $(y,e)$. By maximality, none of them is a tripath. As $x,y\ne1$, this gives $e\in\{1,x\}\cap\{1,y\}$ and $e'\in\{1,y\}\cap\{1,x\}$. As $x\ne y$, this forces $e=e'=1$. Now $ey^{-1}$ and $(e'x^{-1})^{-1}$ are both equal to $[MX_1PY_1^{-1}M^{-1}]$. So $y^{-1}=x$, contradicting Lemma~\ref{lem:no-inverse}.
\end{proof}
 
\begin{lemma}\label{lem:one-block}
If some block of $G$ has non-cyclic image, then $G$ has a rooted spanning tree that is both depth-first and grounded for $f$.
\end{lemma}
 
\begin{proof}
Take a tripath of maximum length, by Lemma~\ref{lem:tripath}. Run a depth-first search from $m$, that follows $M$ to $n$, and then $X$ up to the vertex $x_*$ preceding $m$. Back at $n$, it follows $Y$ up to the vertex $y_*$ preceding $m$. This is possible: otherwise, an inner vertex of $Y$ would be reached from an inner vertex of $X$ through new vertices, avoiding $M$, which is visited first; this contradicts Lemma~\ref{lem:separated}.
 
Let $(T,m)$ be the resulting depth-first rooted spanning tree. The last edges of $X$ and $Y$ are not in $T$. Oriented towards $m$, they emanate from $x_*$ and $y_*$, with labels $x$ and $y$. Oriented away from $m$, they emanate from $m$, with labels $x^{-1}$ and $y^{-1}$. Let $e$ be the first edge of $M$. The side of $e$ containing $m$ is rich, by the second pair of edges. The other side contains $x_*$ and $y_*$, so it is rich by the first pair. So the component $K$ of $T-m$ containing $n$ and $V(G)\setminus K$ are both rich, and every other component $K'$ of $T-m$ has $V(G)\setminus K'\supseteq K$ rich. And $V(G)$, containing $K$, is rich, so $T$ is grounded for $f$.
\end{proof}
 
\begin{proof}[Proof of Proposition~\ref{prop:grounding}]
Some block of $G$ has non-cyclic image or two blocks have non-trivial images by Lemma~\ref{lem:dfs}. Apply Lemma~\ref{lem:one-block} or Lemma~\ref{lem:two-blocks}.
\end{proof}

\paragraph*{AI Statement}

AI was used to assist in writing and generating figures.

\bibliography{refs}

@book{Diestel,
  author    = {Diestel, Reinhard},
  title     = {Graph Theory},
  edition   = {5th},
  series    = {Graduate Texts in Mathematics},
  volume    = {173},
  publisher = {Springer},
  year      = {2017}
}

@book{MoharThomassen,
  author    = {Mohar, Bojan and Thomassen, Carsten},
  title     = {Graphs on Surfaces},
  publisher = {Johns Hopkins University Press},
  year      = {2001}
}

@book{Stillwell,
  author    = {Stillwell, John},
  title     = {Classical Topology and Combinatorial Group Theory},
  edition   = {2nd},
  series    = {Graduate Texts in Mathematics},
  volume    = {72},
  publisher = {Springer},
  year      = {1993}
}

@book{FarbMargalit,
  author    = {Farb, Benson and Margalit, Dan},
  title     = {A Primer on Mapping Class Groups},
  series    = {Princeton Mathematical Series},
  volume    = {49},
  publisher = {Princeton University Press},
  year      = {2012}
}

@article{Kuratowski1930,
  author  = {Kuratowski, Kazimierz},
  title   = {Sur le probl{\`e}me des courbes gauches en topologie},
  journal = {Fundamenta Mathematicae},
  volume  = {15},
  pages   = {271--283},
  year    = {1930},
  doi     = {10.4064/fm-15-1-271-283}
}

@article{Whitney,
  author  = {Whitney, Hassler},
  title   = {A Set of Topological Invariants for Graphs},
  journal = {American Journal of Mathematics},
  volume  = {55},
  number  = {1},
  pages   = {231--235},
  year    = {1933}
}

@article{GloverHunekeWang1979,
  author  = {Glover, Henry H. and Huneke, John Philip and Wang, Chin San},
  title   = {103 Graphs that are Irreducible for the Projective Plane},
  journal = {Journal of Combinatorial Theory, Series B},
  volume  = {27},
  number  = {3},
  pages   = {332--370},
  year    = {1979},
  doi     = {10.1016/0095-8956(79)90022-4}
}

@article{Archdeacon1981,
  author  = {Archdeacon, Dan},
  title   = {A {K}uratowski Theorem for the Projective Plane},
  journal = {Journal of Graph Theory},
  volume  = {5},
  number  = {3},
  pages   = {243--246},
  year    = {1981},
  doi     = {10.1002/jgt.3190050305}
}

@article{RobertsonSeymour1990GM8,
  author  = {Robertson, Neil and Seymour, Paul D.},
  title   = {Graph Minors. {VIII}. {A} {K}uratowski Theorem for General Surfaces},
  journal = {Journal of Combinatorial Theory, Series B},
  volume  = {48},
  number  = {2},
  pages   = {255--288},
  year    = {1990},
  doi     = {10.1016/0095-8956(90)90121-F}
}

@article{HopcroftTarjan1974,
  author  = {Hopcroft, John and Tarjan, Robert},
  title   = {Efficient Planarity Testing},
  journal = {Journal of the ACM},
  volume  = {21},
  number  = {4},
  pages   = {549--568},
  year    = {1974},
  doi     = {10.1145/321850.321852}
}

@article{Williamson1984,
  author  = {Williamson, S. Gill},
  title   = {Depth-First Search and {K}uratowski Subgraphs},
  journal = {Journal of the ACM},
  volume  = {31},
  number  = {4},
  pages   = {681--693},
  year    = {1984},
  doi     = {10.1145/1634.322451}
}

@article{deFraysseixRosenstiehl1985,
  author  = {de Fraysseix, Hubert and Rosenstiehl, Pierre},
  title   = {A Characterization of Planar Graphs by {T}r{\'e}maux Orders},
  journal = {Combinatorica},
  volume  = {5},
  number  = {2},
  pages   = {127--135},
  year    = {1985},
  doi     = {10.1007/BF02579375}
}

@article{Mohar,
  author  = {Mohar, Bojan},
  title   = {Obstructions for the Disk and the Cylinder Embedding Extension Problems},
  journal = {Combinatorics, Probability and Computing},
  volume  = {3},
  number  = {3},
  pages   = {375--406},
  year    = {1994}
}

@article{BlasiusFinkRutter2023,
  author  = {Bl{\"a}sius, Thomas and Fink, Simon D. and Rutter, Ignaz},
  title   = {Synchronized Planarity with Applications to Constrained Planarity Problems},
  journal = {ACM Transactions on Algorithms},
  volume  = {19},
  number  = {4},
  pages   = {34:1--34:23},
  year    = {2023},
  note    = {Preliminary version in ESA 2021}
}

@misc{medina2019plane,
  author = {Medina, Carolina and Mohar, Bojan and Salazar, Gelasio},
  title  = {Well-Quasi-Order of Plane Minors and an Application to Link Diagrams},
  year   = {2019},
  note   = {arXiv:1905.01830},
  url    = {https://arxiv.org/abs/1905.01830}
}

@misc{lunel2025wqo,
  author = {Lunel, Corentin and Maria, Cl{\'e}ment},
  title  = {Well-Quasi-Orders on Embedded Planar Graphs},
  year   = {2025},
  note   = {arXiv:2512.04074},
  url    = {https://arxiv.org/abs/2512.04074}
}

@article{Poincare1904,
  author  = {Poincar{\'e}, Henri},
  title   = {Cinqui{\`e}me compl{\'e}ment {\`a} l'analysis situs},
  journal = {Rendiconti del Circolo Matematico di Palermo},
  volume  = {18},
  pages   = {45--110},
  year    = {1904}
}

@article{Reinhart1962,
  author  = {Reinhart, Bruce L.},
  title   = {Algorithms for {J}ordan Curves on Compact Surfaces},
  journal = {Annals of Mathematics},
  volume  = {75},
  number  = {2},
  pages   = {209--222},
  year    = {1962}
}

@article{Epstein,
  author  = {Epstein, David B. A.},
  title   = {Curves on 2-Manifolds and Isotopies},
  journal = {Acta Mathematica},
  volume  = {115},
  pages   = {83--107},
  year    = {1966}
}

@article{Chillingworth1969,
  author  = {Chillingworth, David R. J.},
  title   = {Simple Closed Curves on Surfaces},
  journal = {Bulletin of the London Mathematical Society},
  volume  = {1},
  number  = {3},
  pages   = {310--314},
  year    = {1969}
}

@article{Chillingworth1972,
  author  = {Chillingworth, David R. J.},
  title   = {Winding Numbers on Surfaces. {II}},
  journal = {Mathematische Annalen},
  volume  = {199},
  pages   = {131--153},
  year    = {1972},
  doi     = {10.1007/BF01431419}
}

@article{FreedmanHassScott1982,
  author  = {Freedman, Michael H. and Hass, Joel and Scott, Peter},
  title   = {Closed Geodesics on Surfaces},
  journal = {Bulletin of the London Mathematical Society},
  volume  = {14},
  number  = {5},
  pages   = {385--391},
  year    = {1982}
}

@article{BirmanSeries1984,
  author  = {Birman, Joan S. and Series, Caroline},
  title   = {An Algorithm for Simple Curves on Surfaces},
  journal = {Journal of the London Mathematical Society (2)},
  volume  = {29},
  number  = {2},
  pages   = {331--342},
  year    = {1984}
}

@article{DespreLazarus2019,
  author  = {Despr{\'e}, Vincent and Lazarus, Francis},
  title   = {Computing the Geometric Intersection Number of Curves},
  journal = {Journal of the ACM},
  volume  = {66},
  number  = {6},
  pages   = {45:1--45:49},
  year    = {2019},
  doi     = {10.1145/3363367}
}

@article{Ladegaillerie1984,
  author  = {Ladegaillerie, Yves},
  title   = {Classes d'isotopie de plongements de 1-complexes dans les surfaces},
  journal = {Topology},
  volume  = {23},
  number  = {3},
  pages   = {303--311},
  year    = {1984}
}

@article{CdVdM2014Isotopy,
  author  = {{Colin de Verdi{\`e}re}, {\'E}ric and de Mesmay, Arnaud},
  title   = {Testing Graph Isotopy on Surfaces},
  journal = {Discrete \& Computational Geometry},
  volume  = {51},
  number  = {1},
  pages   = {171--206},
  year    = {2014},
  doi     = {10.1007/s00454-013-9555-4}
}

@inproceedings{FulekKyncl2018HTApprox,
  author    = {Fulek, Radoslav and Kyn{\v{c}}l, Jan},
  title     = {{H}anani--{T}utte for Approximating Maps of Graphs},
  booktitle = {34th International Symposium on Computational Geometry (SoCG 2018)},
  series    = {LIPIcs},
  volume    = {99},
  pages     = {39:1--39:15},
  year      = {2018},
  doi       = {10.4230/LIPIcs.SoCG.2018.39}
}

@article{AkitayaFulekToth2019,
  author  = {Akitaya, Hugo A. and Fulek, Radoslav and T{\'o}th, Csaba D.},
  title   = {Recognizing Weak Embeddings of Graphs},
  journal = {ACM Transactions on Algorithms},
  volume  = {15},
  number  = {4},
  pages   = {50:1--50:27},
  year    = {2019},
  doi     = {10.1145/3344549}
}

@article{colin2026untangling,
  author  = {{Colin de Verdi{\`e}re}, {\'E}ric and Despr{\'e}, Vincent and Dubois, Lo{\"\i}c},
  title   = {Untangling Graphs on Surfaces},
  journal = {Journal of the ACM},
  volume  = {73},
  number  = {5},
  pages   = {1--46},
  year    = {2026},
  doi     = {10.1145/3830462}
}

\appendix

\section{Loop graphs have small obstructions}\label{app:loops}

We follow~\cite[Section~4]{colin2026untangling}: we lift a straightened drawing of $L$ to the universal cover, where crossings between lifts survive any homotopy. Each such crossing involves at most four loops, whose restriction cannot be untangled; we say that we \emph{return} them.

\begin{proof}[Proof of Lemma~\ref{lem:loops}]
In each component of $L$, remove the loops with contractible images, and merge the loops whose images are homotopic up to orientation. The new map still cannot be untangled, since removed loops could be redrawn as small contractible loops or as parallel copies. If the class of a loop is a proper power, we return that loop, by Lemma~\ref{lem:epstein}.

Assume that $S$ is a sphere, a torus, a disk, or an annulus. Two loops of distinct components with non-zero algebraic intersection number cannot be untangled; we return them. Otherwise, if $L$ is connected, some set of at most four loops cannot be untangled: $L$ itself if it has at most four loops, and otherwise any four of its loops, as by Euler's formula, a one-vertex graph embedded in the torus, with non-contractible loops that are pairwise non-homotopic up to orientation, has at most three loops. If $L$ is not connected, its components are single parallel loops, and $\lambda$ can be untangled.

Otherwise, $S$ has a hyperbolic metric with geodesic boundary. In each component of $L$, call one loop \emph{major} and the others \emph{minor}. Homotope $\lambda$ so that major loops are closed geodesics, minor loops are geodesic loops at their vertex, and distinct vertices have distinct images, by sliding them along the major loops. Two major loops are \emph{parallel} if they have the same image; we orient them alike. A major loop is \emph{pulled} to the left (right) if a minor loop of its component leaves or reaches its vertex from its left (right).

In the universal cover $\tilde S\subseteq\mathbb H^2$, major lifts are geodesic lines, and each minor lift is a geodesic segment joining two lifts of the major loop of its component, which form with it its \emph{H-block}. Two lifts that do not overlap meet at most once. A homotopy of $\lambda$ lifts to an equivariant homotopy of the lifts, which fixes the limit points of the major lifts on $\partial\mathbb H^2$, and the order of the lifted vertices along each major lift.

If two major lifts cross, their limit points interleave, so we return the one or two major loops lifted. So, two major lifts are disjoint or have the same image. If parallel major loops $E\ne E'$ are pulled to the same side by minor loops $e,e'$, or if $E$ is pulled to both sides by $e,e'$, we return $e,e',E,E'$. Indeed, after a homotopy, two lifts of $E$ and $E'$ with the same image become disjoint with the same limit points, so one of them separates the other from the far major lift of the H-block of a lift of $e$ or $e'$.

No minor lift overlaps a major lift $\tilde E$, as the two major lifts of its H-block would then meet $\tilde E$, hence have its image. If a minor lift crosses $\tilde E$, then $\tilde E$ separates the two major lifts of its H-block, and we return the minor loop, the major loop of its component, and the loop lifted by $\tilde E$. So, minor lifts meet major lifts only at their ends.

No two minor lifts overlap, as they would then have the same ends, and lift two loops of the same component homotopic up to orientation. If two minor lifts cross, possibly lifting the same loop, they leave towards the same side any two major lifts of their H-blocks with the same image, which are thus equal, as no two parallel major loops are pulled to the same side. By the order of the lifted vertices, their H-blocks remain crossing, and we return the one or two minor loops and the major loops of their components.

Otherwise, distinct lifts meet only at common ends, except for lifts of parallel major loops. Among parallel major loops, at most one is pulled to the left, at most one to the right, and the others have no minor loop. Perturbing them into disjoint parallel copies, ordered from left to right accordingly, untangles $\lambda$, a contradiction. In all cases, we returned at most four loops.
\end{proof}

\section{Untangling is essentially equivalent to realizing}\label{app:gadget}

\begin{proof}[Proof of Lemma~\ref{lem:gadget}]
Let $N$ be a regular neighborhood of $L$ in $S$: a disk $D$ around the vertex, and one band per loop. We place $C$ on $\partial D$, each vertex at the end of the band of its half-edge. Since $f$ is homotopic to $\mu$, we may replace $f$ by $\mu$.

Assume that $H'$ is realizable, and take a realization of $H'\cup C$ in $D$, with $C=\partial D$. Each pair of matched halves, at $x$ and $\bar x$, reaches the two ends of a band. Join each such pair by an arc through that band. These arcs can be made disjoint because the orders at $x$ and $\bar x$ are reversed. The result is an embedding of $G'$ in $N$. It is homotopic to $\mu|_{G'}$, since edges of $H'-V(C)$ stay in $D$ and the other edges go once along their loop.

Conversely, assume that $\mu|_{G'}$ can be untangled. All edges of $H'-V(C)$ are sent to the vertex. Moreover, $H'-V(C)$ is connected and contains all vertices of $G'$. So $H'-V(C)$ contains a spanning tree of $G'$ sent to the vertex; up to deleting unused loops of $L$, $\lambda\circ\mu|_{G'}$ is the factorization of $f|_{G'}$ along it, and Lemma~\ref{lem:factorization-untangle} shows that $\mu|_{G'}$ is a weak embedding. Hence some embedding of $G'$ close to $\mu|_{G'}$ lies in $N$, with its vertices in $D$. Cut each band along a cocore. Removing bigons between the edges and the cocore inside the band, we may assume that each edge sent around the loop crosses it once and that the other edges avoid it. Cutting the bands along their cocores and retracting them into $D$ gives a realization of $H'$. The orders at matched vertices are reversed because $S$ is orientable.
\end{proof}

\section{Richness for a map and for its gadget coincide}\label{app:transfer}

\begin{proof}[Proof of Lemma~\ref{lem:transfer}]
A homotopy conjugates all labels by the same element, so we may compute labels with $\lambda\circ\mu$. An edge sent to the vertex has trivial label. An edge sent around a loop has, as label, the class of that loop with the orientation in which the edge traverses it. As homotopic loops were bundled and contractible ones removed, distinct oriented loops have distinct non-trivial classes. This also holds for a loop and its inverse, as $\pi_1(S)$ has no torsion.

For an oriented edge sent around a loop, the oriented loop corresponds to the vertex of $C$ of the half at its tail. So two oriented edges emanating from $U$ have distinct non-trivial labels if and only if their halves at $U$ lie at distinct vertices of $C$. In other words, $U$ is rich for $f$ if and only if it is rich in $(H,C)$. So $(T,r)$, being grounded for $f$, is grounded in $(H,C)$.
\end{proof}

\section{Cyclic maps have small obstructions}\label{app:cyclic}

This section proves Lemma~\ref{lem:annular}. We look for two kinds of obstructions: three non-contractible cycles that cannot be nested, and a contractible cycle forced to hold a non-contractible piece. We find them by peeling the graph: cut vertices, then separation pairs, then a single $3$-connected node. Routine verifications are left to the reader.

Throughout, $G$ is connected and $f:G\to S$ is cyclic and \emph{critical}: it cannot be untangled, but its restriction to every proper subgraph can. A subgraph is \emph{trivial} if its image is. A graph is \emph{small} if its cycle rank is at most nine. The \emph{winding} of a closed walk in an open annulus $A$ is its class in $\pi_1(A)\cong\mathbb Z$.

\subsection{The annulus}

\begin{lemma}\label{lem:setup}
If $G$ is not small, we may assume that $f$ maps $G$ into an open annulus, that $G$ is planar and not trivial, and that cycles have winding $0$ or $\pm1$.
\end{lemma}

\begin{proof}
If $f$ is trivial, or $G$ is not planar, $G$ is a Kuratowski subgraph by criticality. If a cycle cannot be untangled, $G$ is that cycle. Otherwise, by Lemma~\ref{lem:epstein}, the image of $f$ is generated by the class of a simple closed curve $\delta$, and cycles have classes $\delta^k$ with $k\in\{0,\pm1\}$, up to conjugacy. Lift $f$ to the annular cover associated with $\langle\delta\rangle$: lifts of embeddings are embeddings, and an embedding in the cover can be pushed near the lift of $\delta$, where the covering map is injective.
\end{proof}

From now on, $f$ maps $G$ into $A$, the sphere minus two \emph{ends} $e^\pm$, oriented so that a simple closed curve with $e^+$ on its left and $e^-$ on its right has winding $1$. Facial walks are oriented with their face on their left.

\begin{lemma}\label{lem:faces}
Let $K\subseteq G$ be connected. An embedding of $K$ in the sphere minus $e^\pm$ is homotopic to $f|_K$ if and only if each face $F$ has a facial walk of winding $[e^+\in F]-[e^-\in F]$.
\end{lemma}

\begin{proof}
The facial walk of $F$ has this winding, facial walks generate the cycle space, and maps to $A\simeq\mathbb S^1$ are determined up to homotopy by the windings of cycles.
\end{proof}

If $K$ is not trivial, the ends lie in distinct faces, the \emph{annular} faces, whose facial walks contain non-contractible cycles; a vertex on them is \emph{exposed}. Mirroring an embedding lets us choose which end lies in which annular face. Finally, a face of a connected $K$ containing a non-contractible simple closed curve contains an end, as $K$ lies on one side of the curve.

\subsection{Two obstructions}

Three non-contractible cycles $O_1,O_2,O_3$, pairwise meeting in connected, possibly empty, sets, form a \emph{nested triple} if $G-V(O_b)$ joins $O_a$ to $O_c$ for all $\{a,b,c\}=\{1,2,3\}$. A \emph{split} is a contractible cycle $O$ with two interlacing bridges, each forming a non-contractible cycle with $O$.

\begin{lemma}\label{lem:nesting}
If $G$ has a nested triple or a split, then $G$ is small.
\end{lemma}

\begin{proof}
In an embedding homotopic to $f$, the sides of the $O_i$ containing $e^+$ are nested, so the middle one separates the other two. This persists in the union of the three cycles, of cycle rank at most four, and of at most three paths joining them as required; this graph is small, and is $G$ by criticality.

In an embedding, the contractible cycle $O$ of a split bounds a disk without ends, which contains neither bridge, so the bridges lie on the same side of $O$ and cross. Each bridge has a path forming a non-contractible cycle with an arc of $O$, as windings add up. Keeping $O$ and, in each bridge, a tree with at most five leaves containing this path and attachments realizing the interlacing gives a graph of cycle rank at most $1+4+4$, which is $G$.
\end{proof}

\subsection{Cut vertices}

\begin{lemma}\label{lem:trivial parts}
Let $G=G_1\cup G_2$, with no common edge, and $G_2$ connected, trivial, with an edge. Then $G_1\cap G_2$ is not a vertex; and if it is $\{u,v\}$, and $G_1$ has a $u$--$v$ path, then $G_2$ is a $u$--$v$ path.
\end{lemma}

\begin{proof}
Otherwise, untangle $f$ on $G_1$, or on $G_1$ plus a $u$--$v$ path $P\subsetneq G_2$, and redraw $G_2$ in a small disk, or in a thin disk along $P$; $G_2+uv$ is planar as $G$ is. This contradicts criticality.
\end{proof}

A \emph{pendant} at $x$ is a non-contractible cycle through $x$, or a \emph{lollipop}: a non-contractible cycle avoiding $x$ with a path from $x$ to it. In the next three lemmas, $G_1\cap G_2=\{x\}$, and $G_1,G_2$ are connected and not trivial.

\begin{lemma}\label{lem:vertex gluing}
If $f$ can be untangled on $G_1$ and on $G_2$, each time with $x$ exposed, then it can be on $G_1\cup G_2$.
\end{lemma}

\begin{proof}
Up to mirror image, put $G_2$ in a punctured disk bounded by a loop at $x$ in an annular face of $G_1$.
\end{proof}

\begin{lemma}\label{lem:light gluing}
If $x$ lies on every non-contractible cycle of $G_2$, and $f$ can be untangled on $G_2$ and on $G_1\cup O$ for a non-contractible cycle $O\subseteq G_2$, then it can be on $G_1\cup G_2$.
\end{lemma}

\begin{proof}
The annular faces of $G_2$ have facial walks through $x$, so $G_2$ lies between two loops at $x$; put it in a thin neighborhood of $O$.
\end{proof}

\begin{lemma}\label{lem:exposed}
If $f$ can be untangled on $G_1\cup G_2$, and $G_2$ contains a lollipop at $x$, then it can be on $G_1$ with $x$ exposed.
\end{lemma}

\begin{proof}
The lollipop lies in a face of $G_1$ containing a non-contractible cycle, which is annular.
\end{proof}

\begin{lemma}\label{lem:cut vertices}
If $G$ is not small, it consists of a block $H$ and at most two pendants, at distinct vertices of $H$, the \emph{marks}, each meeting $H$ only there.
\end{lemma}

\begin{proof}
Let $x$ be a cut vertex, $Y$ a component of $G-x$ with $x$, and $Y'$ the rest; both are non-trivial by Lemma~\ref{lem:trivial parts}. One of them is a pendant: if $x$ lies on every non-contractible cycle of $Y$, then $Y$ is a cycle, by Lemma~\ref{lem:light gluing} and criticality; likewise for $Y'$; otherwise, they contain lollipops $L,L'$ at $x$, and $Y=L$ or $Y'=L'$, by Lemmas~\ref{lem:exposed} and~\ref{lem:vertex gluing} applied to $Y\cup L'$ and $Y'\cup L$. If all components at $x$ give pendants, three of them are lollipops, by Lemma~\ref{lem:vertex gluing}, and form a nested triple. Otherwise, each cut vertex $x$ has exactly one pendant $P_x$, and these pendants are nested along the block--cut tree; removing the maximal ones leaves a block $H$. Three of them would form a nested triple through $H$.
\end{proof}

\begin{lemma}\label{lem:marks}
Let $K\subseteq H$ be connected and not trivial. Then $f$ can be untangled on $K$ plus the pendants at some marks of $K$ if and only if it can be on $K$ with these marks exposed, on distinct annular faces.
\end{lemma}

\begin{proof}
Each pendant lies in an annular face at its mark, and two pendants in one annular face would cross. Conversely, apply Lemma~\ref{lem:vertex gluing}.
\end{proof}

\begin{lemma}\label{lem:H trivial}
If $G$ is not small, then $H$ is not trivial.
\end{lemma}

\begin{proof}
If $H$ is trivial, $f$ can be untangled if and only if $H$, plus an edge $e$ between the marks, is planar. Otherwise, a Kuratowski subgraph of $H+e$, minus $e$, with the pendants, has cycle rank at most $5+2$, and $f$ cannot be untangled on it.
\end{proof}

\subsection{Separation pairs}

From now on, $G$ is not small. If $H$ is a cycle, $G$ is small; otherwise, we suppress the vertices of degree two of $H$ other than the marks, and consider the decomposition of $H$ into $3$-connected components, a tree $\mathcal T$ whose nodes are cycles (S), bonds (P), and $3$-connected simple graphs (R). Each tree edge, with virtual edges $uv$, splits $H$ into two \emph{sides} $X$ and $Y$, with $X\cap Y=\{u,v\}$, $X+uv$ and $Y+uv$ $2$-connected, and at least two edges each. A side $X$ is a \emph{branch} if $X-\{u,v\}$ is connected, which fails only for the side containing an incident P-node. A side \emph{carries} the marks in $V(X)\setminus\{u,v\}$. For $u$--$v$ walks $p,q$, $w(p,q)$ is the winding of $pq^{-1}$.

A \emph{probe} of a non-trivial side $X$ is a non-contractible cycle $O$ of $X$, avoiding $u$ or $v$ if possible, plus disjoint paths from $u$ and $v$ to $O$; its \emph{routes} are its two $u$--$v$ paths. A side is \emph{small} if it is (S1) a $u$--$v$ path through a single mark, (S2) a probe carrying no mark, or (S3) a \emph{bundle}: two paths of type (S1), $p,q$ with $w(p,q)=0$.

\begin{lemma}\label{lem:thin}
In a non-trivial branch, some non-contractible cycle avoids $u$ or $v$.
\end{lemma}

\begin{proof}
Otherwise, in an untangling, $u$ and $v$ lie on both annular faces, and two arcs joining them through these faces would separate $X-\{u,v\}$ from a path between the ends avoiding $X$.
\end{proof}

\begin{lemma}\label{lem:sides}
A trivial branch $X$ is of type \textup{(S1)}.
\end{lemma}

\begin{proof}
If $X$ carries no mark, it is an edge by Lemma~\ref{lem:trivial parts}, not a side. If it carries two, the other side $Y$ is not trivial, for the same reason, and its non-contractible cycles form nested triples with the two pendants. So $X$ carries one mark $x$, with pendant $P$. Let $Z$ be a probe of $Y$, or a $u$--$v$ path of $Y$ if $Y$ is trivial. In an untangling of $X\cup Z\cup P$, the ends lie in one face of $X$, which contains $P$ and, as non-contractible cycles cross it through $Z$, has $u$ and $v$ on its boundary. So $X$ plus a vertex $w$ adjacent to $u,x,v$ is planar: otherwise, a Kuratowski subgraph of it, with $w$ replaced by $P$ and $Z$, has cycle rank at most $5+1+2$, and is $G$. Then $X$ has a $u$--$v$ path $P'$ through $x$, and untangling $f$ with $X$ replaced by $P'$, then redrawing $X$ along $P'$ away from $P$, shows that $X=P'$ by criticality.
\end{proof}

Let $X,Y$ be non-trivial sides, with probes $Q_X,Q_Y$. Let $G_X$ consist of $X\cup Q_Y$, the pendants carried by $X$, and those at $u$ and $v$; define $G_Y$ likewise.

\begin{lemma}\label{lem:two vertices}
If $X$ is a branch, and $f$ can be untangled on $G_X$ and $G_Y$, then it can be on $G$.
\end{lemma}

\begin{proof}
In an untangling of $G_X$, $Q_Y$ lies in an annular face $F_X$ of $X$, by Lemma~\ref{lem:thin} if $Q_Y$ is a cycle through $u$ and $v$. As $X+uv$ is $2$-connected, $F_X$ is bounded by two $u$--$v$ paths $A_1,A_2$ of $X$, with $w(A_1,A_2)=\pm1$. Likewise for $Y$, with $B_1,B_2$. As cycles have winding at most one, $\{w(A_1,R),w(A_2,R)\}=\{w(B_1,R),w(B_2,R)\}$ for a fixed path $R$. So we can glue $X$ and $Y$, using the mirror image of $Y$ if needed, so that $A_1\cup B_j$ and $A_2\cup B_{3-j}$ bound faces of winding $0$; by Lemma~\ref{lem:faces}, this untangles $f$ on $H$, with the other annular faces $F'_X$ and $F'_Y$ of $X$ and $Y$. A mark carried by $X$ lies on $F'_X$; so does a mark at $u$ if every non-contractible cycle of $X$ contains $u$, and likewise for $Y$. Checking the possible pairs of marks, either the marks lie on distinct faces among $F'_X,F'_Y$, and Lemma~\ref{lem:marks} concludes, or the pendants and cycles of $X$ or $Y$ avoiding a mark form a nested triple.
\end{proof}

\begin{lemma}\label{lem:node}
Some R-node $\mu$ has only small sides beyond it, that is, for each tree edge at $\mu$, the side not containing $\mu$ is small.
\end{lemma}

\begin{proof}
Let $X,Y$ be the sides of a tree edge, $X$ a branch. If $X$ is trivial, it is small by Lemma~\ref{lem:sides}. If $X,Y$ are not trivial, then $X=Q_X$ or $Y=Q_Y$, carrying no mark, by Lemma~\ref{lem:two vertices} and criticality. If only $Y$ is trivial, it is a branch, small by Lemma~\ref{lem:sides}, or consists of the other branches and real edges at a P-node, which are paths of type (S1) with the same winding; by criticality, they form a bundle. If both sides are small, $G$ is small. Otherwise, orienting each tree edge towards the node beyond which its side is small, we get a node $\mu$ with only small sides beyond it. If $\mu$ is a cycle, it has at most two probes, as three form a nested triple, and $G$ has cycle rank at most $1+2+1+2$. If $\mu$ is a bond, it has at most two real edges, as equal windings contradict criticality, at most two probes, as three cannot be untangled, and at most two paths of type (S1); so $G$ has cycle rank at most $5+2+2$.
\end{proof}

\subsection{A \texorpdfstring{$3$}{3}-connected node}

Let $\mu$ be as in Lemma~\ref{lem:node}. Its embedding is unique up to mirror image~\cite{Whitney}, its facial cycles are induced and non-separating, and two faces meet in at most an edge. The embeddings of $H$ are obtained by drawing each probe or bundle at its virtual edge $uv$ in one of two ways. Besides the faces of $\mu$, modified at probes and bundles, this creates a face of winding $0$ in each bundle, and a face $D_Q$ of winding $\pm1$ inside the cycle $O_Q$ of each probe $Q$. A face of $\mu$ is \emph{free} if it is not at a probe. The \emph{lens} of a probe at $uv$ is the cycle formed by the two $u$--$v$ paths $\Lambda_1,\Lambda_2$ bounding the faces $f_1,f_2$ at $uv$.

\begin{lemma}\label{lem:rigid triple}
Three non-contractible cycles, each a free facial cycle, a cycle $O_Q$, or the cycle of a pendant whose mark is on neither other, form a nested triple.
\end{lemma}

\begin{proof}
They pairwise meet in at most an edge, and $\mu$ minus a facial cycle, or minus two vertices, is connected.
\end{proof}

\begin{lemma}\label{lem:lens}
If the lens of a probe $Q$ is contractible, $Q$ and any other bridge of the lens containing a non-contractible cycle form a split.
\end{lemma}

\begin{proof}
By planarity and $3$-connectivity, every bridge of the lens in $\mu$ other than $uv$ attaches inside both $\Lambda_1$ and $\Lambda_2$.
\end{proof}

Let $R,R'$ be the routes of a probe $Q$ at $uv$, with $w(R',R)=1$, orient $uv$ with $f_1$ on its left, and let $s=w(R,\Lambda_1)$ and $t=w(\Lambda_2,R)$. Then $s\in\{-1,0\}$ and $t\in\{0,1\}$, the lens has winding $\ell=s+t$, and the faces derived from $f_1,f_2$ and $D_Q$ have windings $(s,t-1,1)$ if $R$ bounds $f_1$, and $(s+1,t,-1)$ otherwise.

\begin{lemma}\label{lem:rigid}
In the setting of Lemma~\ref{lem:node}, $G$ is small.
\end{lemma}

\begin{proof}
We untangle $f$, or find a nested triple or a split. By Lemma~\ref{lem:rigid triple}, there are at most two probes. A bundle carries both marks, and with a probe $Q$ their pendants and $O_Q$ form a nested triple.

\emph{No probe.} Exactly two faces of $\mu$, $F$ and $F'$, have non-zero winding, or three form a nested triple. A mark on neither, or two marks only on $F$, give a nested triple; with a bundle, so does a face $F$ not at it. Otherwise, the marks can be placed.

\emph{One probe $Q$.} At most one free face $F_0$ has non-zero winding. If there is one, then $\ell=-w(F_0)$, so $(s,t)$ is $(0,1)$ or $(-1,0)$, and one drawing of $Q$ makes the faces at $uv$ zero; the annular faces are $F_0$ and $D_Q$, and marks are handled as before. If there is none, then $\ell=0$, and the two drawings of $Q$ leave, besides $D_Q$, one non-zero face, derived from $f_1$ or $f_2$ respectively. A mark off the lens and $O_Q$ gives a split by Lemma~\ref{lem:lens}, and two marks off $O_Q$ give a nested triple; otherwise, the marks can be placed.

\emph{Two probes $Q_1,Q_2$.} Free faces have winding $0$. If the lenses share no face and are contractible, $Q_1$ and the bridge of its lens containing $Q_2$ form a split. Otherwise, we draw the probes so that only $D_{Q_1}$ and $D_{Q_2}$ are non-zero: as in the previous case if the lenses share no face; and if they share a face, by making the other two faces zero, which forces the shared one to be zero, its winding being at most one. A mark on neither $O_{Q_1}$ nor $O_{Q_2}$, or two marks only on one of them, give a nested triple; otherwise, the marks can be placed.
\end{proof}

\begin{proof}[Proof of Lemma~\ref{lem:annular}]
Take $G'\subseteq G$ minimal such that $f|_{G'}$ cannot be untangled. It is connected, as untangled components can be nested in the annular cover, so we may assume that $f$ is critical, and it suffices to show that $G$ is small; then $G$ has minimum degree at least two by Section~\ref{sec:prelim}, so it has at most $2\cdot9-2=16$ topological vertices. If $G$ is not small, Lemmas~\ref{lem:setup}, \ref{lem:cut vertices} and~\ref{lem:H trivial}, then Lemma~\ref{lem:node}, reduce it to a single R-node, and Lemma~\ref{lem:rigid} gives a contradiction.
\end{proof}

\section{Appendix of Section~\ref{sec:core}}\label{app:core}

\begin{proof}[Proof of Lemma~\ref{lem:merge}]
While two vertices of $F$ on $U$ lie in distinct components of $F-V(C)$, the path of $U$ between them, which avoids $C$, has a subpath with ends in distinct components of $F-V(C)$ and no inner vertex in $F$. Adding it to $F$ merges two of these components, so the cycle rank of $F-V(C)$ does not change.
\end{proof}

\end{document}